\documentclass[aps,prb,10pt,twocolumn,superscriptaddress,notitlepage,floatfix,bibnotes]{revtex4-2}
\usepackage{amsmath,amsfonts,amssymb,bm}
\usepackage{hyperref,graphicx}
\usepackage{xcolor,xspace}

\newcommand{\B}{\mathcal{B}}
\newcommand{\C}{\mathcal{C}}
\newcommand{\V}{\mathcal{V}}
\newcommand{\F}{\mathbb{F}}
\newcommand{\I}{\mathbb{I}}

\newcommand{\Z}{\mathbb{Z}}

\DeclareMathOperator{\rank}{\mathrm{rank}}
\DeclareMathOperator{\img}{\mathrm{img}}

\begin{document}

\title{Fracton models from product codes}

\date{\today}

\author{Yi Tan}
\thanks{These authors contributed equally to this work.}
\affiliation{Department of Physics, Harvard University, Cambridge, MA 02138, USA}

\author{Brenden Roberts}
\thanks{These authors contributed equally to this work.}
\affiliation{Department of Physics, Harvard University, Cambridge, MA 02138, USA}

\author{Nathanan Tantivasadakarn}
\affiliation{Walter Burke Institute for Theoretical Physics and Department of Physics, California Institute of Technology, Pasadena, CA 91125, USA}

\author{Beni Yoshida}
\affiliation{Perimeter Institute for Theoretical Physics, Waterloo, Ontario N2L 2Y5, Canada}

\author{Norman Y.~Yao}
\affiliation{Department of Physics, Harvard University, Cambridge, MA 02138, USA}
  
\begin{abstract}
We explore a deep connection between fracton order and product codes.
In particular, we propose and analyze conditions on classical seed codes which lead to fracton order in the resulting quantum product codes. 
Depending on the properties of the input codes, product codes can realize either Type-I or Type-II fracton models, in both nonlocal and local constructions.
For the nonlocal case, we show that a recently proposed model of lineons on an irregular graph can be obtained as a hypergraph product code. 
Interestingly, constrained mobility in this model arises only from glassiness associated with the graph.
For the local case, we introduce a novel type of classical LDPC code defined on a planar aperiodic tiling. 
By considering the specific example of the pinwheel tiling, we demonstrate the systematic construction of local Type-I and Type-II fracton models as product codes. 
Our work establishes product codes as a natural setting for exploring fracton order.
\end{abstract}
\maketitle

Recent years have seen ideas and techniques from quantum information drive new developments in many areas of physics.
Among these are new insights around an old problem in statistical mechanics, namely, the physics of glassiness.
Exotic models with glassy phenomenology typify both the emerging field of fracton topological order~\cite{Chamon_PRL2005,Haah_PRA2011,Yoshida_PRB2013,Vijay_PRB2015,Vijay_PRB2016,Slagle_PRB2017,Pretko_PRB2018,Gromov_PRX2019,Shirley_Foliated2019,Dua_Sorting2019,Nandkishore_ARCMP2019,Pretko_IJMPA2020} as well as breakthrough developments of asymptotically good quantum low-density parity check (LDPC) codes implemented in non-local stabilizer Hamiltonians~\cite{Breuckmann_TIT2021,Panteleev_ACM2022, Breuckmann_PRX2021}.

\begin{figure}[ht!]
\centering
\includegraphics[width=\columnwidth]{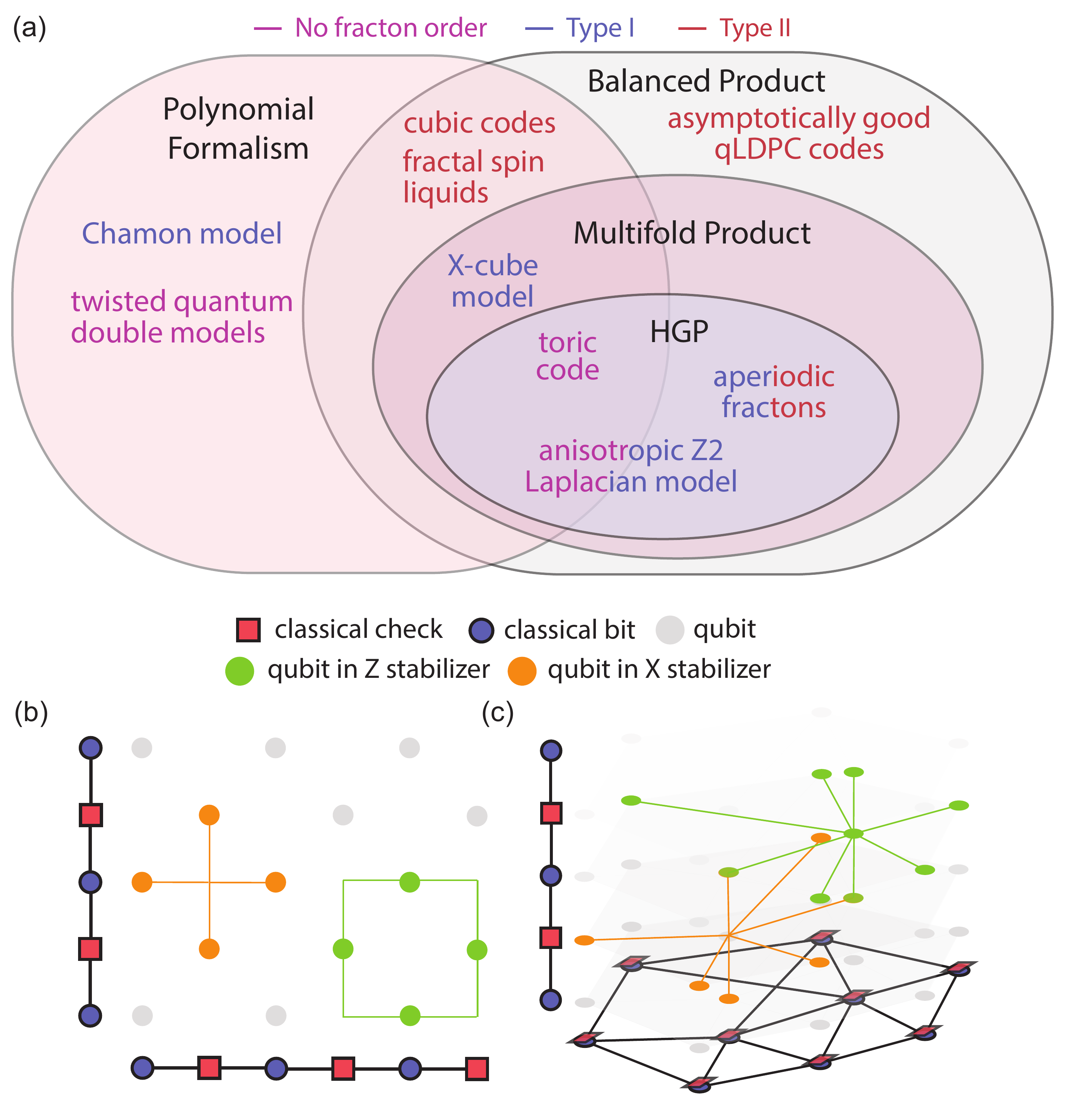}
\caption{\label{fig:venn} (a) Set inclusions are shown for stabilizer models.
Some explicit product code constructions are provided in the supplementary material~\cite{Supplemental}.
(b,c) Classical Tanner graphs are depicted, with bits represented by blue circles and checks by red squares.
The HGP is represented graphically: in (b), a surface code is found as the HGP of two repetition codes, and in (c), the anisotropic $\Z_2$ Laplacian model~\cite{Gorantla_PRB2023,Ebisu_Scipost2023} is obtained from a repetition code and a Laplacian code.
}
\end{figure}

At first glance, these two topics seem rather distinct. 
On the one hand, fracton phases of matter constitute a fascinating counterexample to the lore of solid state physics by reproducing glassy dynamics within local models exhibiting translation invariance~\cite{Chamon_PRL2005}.
Features distinguishing such phases from both symmetry breaking and conventional topological order include: (i) ground-state degeneracy in finite systems that depends nontrivially on the volume, and (ii) low-energy excitations with constrained or no mobility.
Instances with constrained mobility are referred to as Type-I fracton models, and those with no mobility as Type-II fracton models~\cite{Haah_PRA2011, Yoshida_PRB2013, Vijay_PRB2015, Vijay_PRB2016}.

On the other hand, quantum LDPC codes represent a promising type of quantum error correction that can be defined via a stabilizer Hamiltonian $\mathcal{H}=\sum_j \Pi_j$, where each term $\Pi_j$ has bounded norm and acts on a constant number of qubits, and each qubit participates in only a constant number of terms.
A systematic approach to generating such codes is the family of product constructions, which take as input multiple classical LDPC codes and output a quantum LDPC code~\cite{Hastings_arxiv2021,Panteleev_ITI2022,Breuckmann_TIT2021}.
The canonical example is the \emph{hypergraph product} (HGP), a prescription for generating a CSS quantum code from a pair of classical codes~\cite{Tillich_TIT2014}.
Crucially, product codes allow analytic calculation of the resulting code parameters including the encoding rate (the number of logical qubits per physical qubit) and the code distance (the minimal number of errors required to connect logical states).  

Despite their different contexts, these topics are connected by certain individual results: for instance, Haah's code, a Type-II fracton model, can be obtained via a generalization of the HGP known as the \emph{lifted product}~\cite{Panteleev_ITI2022,Panteleev_ACM2022}.
One can also use classical codes to construct quantum ``fractal spin liquids''~\cite{Yoshida_PRB2013}, based on the polynomial formalism for translation invariant codes~\cite{Haah_Commuting2013,*Haah_Algebraic2016}~(Fig.~\ref{fig:venn}).

More broadly, an interesting observation suggests a fundamental connection between fracton order and HGP codes.
To see this, we note that, for gapped phases of matter, the logarithm of the ground-state degeneracy determines the number of qubits that could in principle be encoded in the ground space. 
In conventional phases, this effective qubit number is independent of system size.
This scaling is evidently incompatible with the number of logical qubits in typical HGP codes, which is polynomial in the system size.
By contrast, such polynomial scaling is reminiscent of precisely the nontrivial dependence of the ground-state degeneracy on system size in fracton phases. 
Nevertheless, a sharp understanding of the relationship between fracton order and product codes remains lacking.

In this Letter, we provide such a framework, and demonstrate that product codes provide a natural setting for discovering new fracton phases of matter (Fig.~\ref{fig:venn}). 
Our main results are threefold.
First, we extend the definition of fracton order to stabilizer models without translation invariance or geometric locality.
Based on this definition,  we propose a set of criteria on the input classical codes of a product construction.
When satisfied, the resulting quantum LDPC code realizes a nonlocal fracton phase.
Second, we show that a recent model, purported to exhibit Type-I fracton order on an irregular graph, can be obtained as a HGP code. 
Interestingly, while this case does not, constitute a fracton model, a generalized construction (utilizing a typical LDPC input code) does.

\begin{figure}
\centering
\includegraphics[width=\columnwidth]{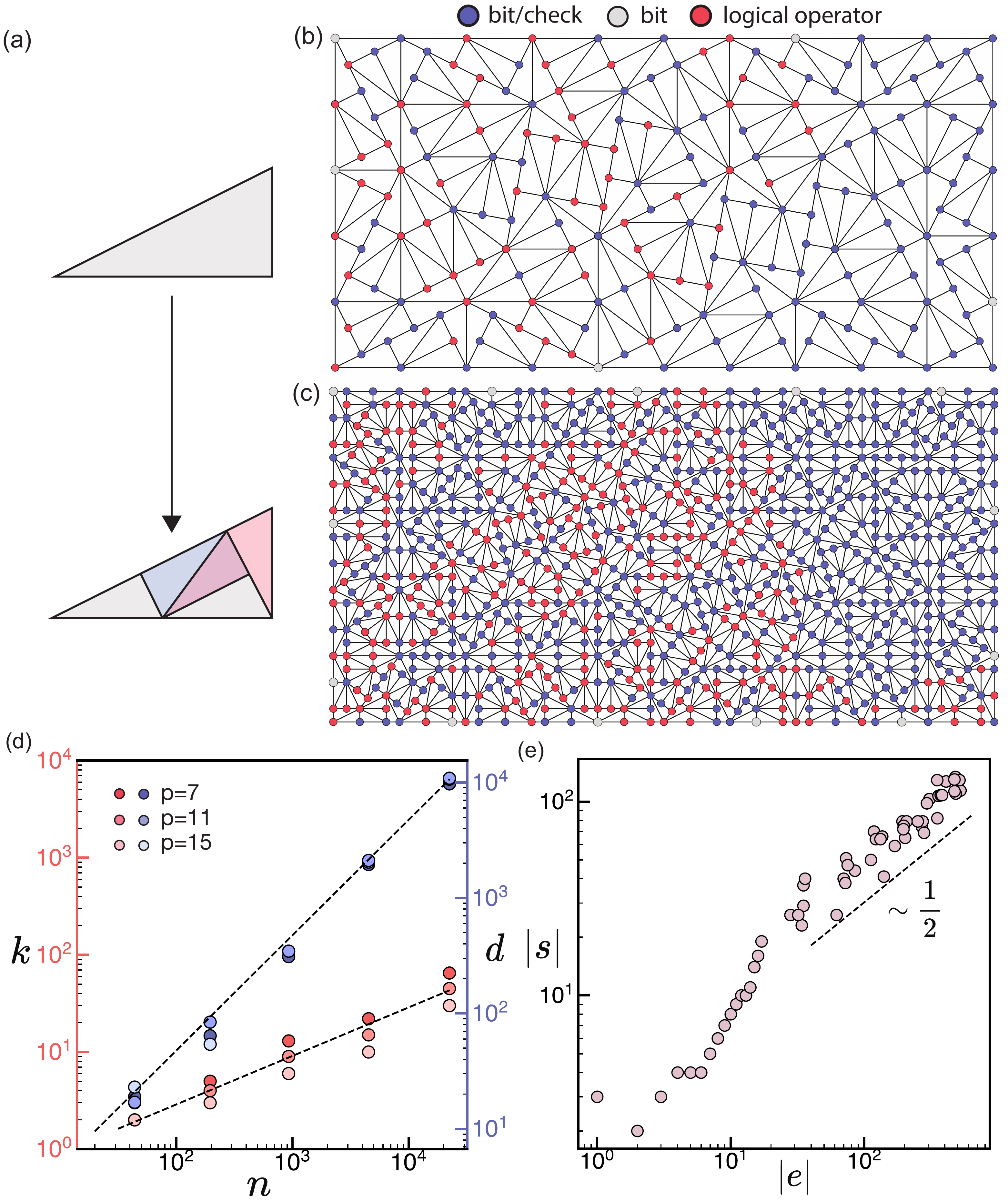}
\caption{\label{fig:pinwheel_code}
(a) The pinwheel tiling substitution rule is shown.
(b,c) Graphs $G_N$ at generations $N=3,4$ are shown for the pinwheel tiling on a rectangle.
Each vertex hosts both a bit and a check, except for a fraction of boundary sites, whose checks are removed.
The lowest-weight codeword is indicated in red.
(d) The scaling of the number of logical bits $k \sim \sqrt n$ and distance $d \sim n$ are shown for the pinwheel code, for $p=7,11,15$.
Distances shown are exact, except for the largest value $k = 65$, for which an upper bound is shown.
(e) The minimum syndrome weight over sampled errors at $N=5$ provides numerical evidence for confinement.
For low-weight errors, all error configurations are considered within randomly sampled balls of fixed radius.
For high-weight errors, truncated logical operators illustrate the naive square-root upper bound.
Though not exhaustive, such sampling can diagnose lack of confinement in practice~\cite{Supplemental}.}
\end{figure}

Finally, we consider \emph{geometrically local} product constructions, introducing a new family of classical LDPC codes defined on aperiodic tilings of the plane.
Specifically, we consider a local two-dimensional code on the pinwheel tiling (Fig.~\ref{fig:pinwheel_code}), which we term the \emph{pinwheel code}.
We show that the HGP of the pinwheel code with a repetition code produces a local Type-I fracton model in three dimensions, and that the HGP of two pinwheel codes produces a local Type-II fracton model in four dimensions. 
These results open the door to the systematic discovery of new \emph{local} fracton models using product codes.
The resulting models also exhibit evidence of a feature known as confinement (see discussions below), a favorable property the is lacking from most known fracton models~\cite{Bombin_PRX2015,Quintavalle_PRX2021}.

\emph{Hypergraph product codes.---}A classical LDPC code on $n$ bits with $m$ terms is represented by a $\kappa$-sparse $m \times n$ parity check matrix $H \in \F_2^{m\times n}$, where $\F_2 = \{0,1\}$.
It is characterized by a triplet $[n,k,d]$, where $k$ is the number of encoded logical bits and $d$ is the code distance~\cite{MacKay_2003}.
The number of bits participating in each check is bounded above by a constant $\kappa_c$, and the number of checks acting on any given bit is bounded  by $\kappa_b$; then $\kappa = \max(\kappa_b,\kappa_c) = O(1)$.
Typical instances of LDPC codes are obtained by randomly sampling parity check matrices $H$~\cite{Supplemental}.

Similarly, a CSS quantum code is represented by two parity check matrices $H_X$ and $H_Z$ specifying the action of the $X$ and $Z$ stabilizers, respectively~\cite{Calderbank_PRA1996,Steane_PRL1996}.
Both classical and quantum parity check matrices are maps taking a configuration of errors in $\F_2^n$, to a syndrome or excitation in $\F_2^m$.

The HGP acts on a pair of classical ``seed'' codes, producing a CSS quantum code which is LDPC if the inputs are LDPC~\cite{Tillich_TIT2014}.
The quantum code resulting from the HGP of classical seed codes $H_1\in \F_2^{m_1\times n_1}$ and $H_2\in\F_2^{m_2\times n_2}$ is given by:
\begin{equation}
H_X = \begin{bmatrix}H_1\otimes \I & \I\otimes H_2^\top\end{bmatrix},~
H_Z = \begin{bmatrix}\I\otimes H_2 & H_1^\top\otimes\I\end{bmatrix},
\label{eq:hgp}
\end{equation}
with $\I$ being the identity matrix.
Classical seed codes with parameters $[n_1,k_1,d_1]$ and $[n_2,k_2,d_2]$ produce a quantum code with parameters $[[n_q,k_q,d_q]]$, where~\cite{Tillich_TIT2014}
\begin{equation}
\renewcommand*{\arraystretch}{1.4}
\begin{matrix}
n_q = n_1 n_2 + m_1 m_2~,~d_q = \min(d_1,d_2,d_1^\top,d_2^\top), \\
k_q = k_1 k_2 + k_1^\top k_2^\top~,~k_X^\top = k_1^\top k_2~,~k_Z^\top = k_1 k_2^\top~.
\end{matrix}
\label{eq:hgp_params}
\end{equation}
Here, $k_i^\top$ and $d_i^\top$ denote the parameters of the transposed (or dual) code, with parity check matrix $H_i^\top$.
The number of \emph{superselection sectors} in the quantum code is given by $2^{k_X^\top + k_Z^\top}$~\cite{Supplemental}.
The prototypical example of a HGP is the toric code, obtained from two cyclic repetition codes having $k_1 = k_1^\top = k_2 = k_2^\top = 1$ and $d_1 = n_1$, $d_2 = n_2$~\cite{Tillich_TIT2014}.

As shown in Fig.~\ref{fig:venn}(b), the HGP admits an intuitive graphical representation based on the \emph{Tanner graph}; a bipartite Tanner graph describes any classical code by representing both checks and bits as vertices, with edges denoting the connectivity of the terms in $H$~\cite{Tanner_IEEE1981}.

\emph{Fracton order in nonlocal product codes}.---Product constructions of stabilizer codes generically do not feature spatial symmetry or geometric locality; to connect to  conventional fracton models~\cite{Nandkishore_ARCMP2019,Pretko_IJMPA2020} and the local codes that we discuss later, we begin by providing a concrete definition of nonlocal fracton order.
In particular, we sample from a random ensemble specifying a family of codes with connected Tanner graphs as $n_q\to\infty$.
Such a family is said to host fractons if: the logarithm of the ground state degeneracy scales polynomially in the system size~\footnote{Note that this scaling is distinct from that of local fractons, which merely follows a polynomial envelope.}; and excitations are pointlike \footnote{While finite-dimensional local fracton models with immobile loop excitations are known~\cite{LiYe20}, such a case is not expected for the nonlocal models we consider.} and have constrained mobility due to superselection sectors which also scale polynomially in the system size.
An additional requirement in the case of nonlocal Type-II fractons is ``confinement''~\cite{Quintavalle_PRX2021}, a natural consequence of nonlocality that implies the absence of string-like operators~\footnote{The definition of confinement here is weaker than its usage elsewhere~\cite{Bombin_PRX2015}, as the energy cost is allowed to scale as any increasing function of the error weight, rather than strictly linearly.}.
The essence of confinement is that for sparse error patterns, additional errors always increase syndrome weight.

Product constructions provide the opportunity to reason about quantum codes through investigation of classical seeds.
We propose three criteria for classical codes leading to nonlocal fracton order in an HGP code: \emph{rank deficiency}, \emph{confinement}, and \emph{isolability}.

\emph{Rank deficiency} refers to a property of the parity check matrix $H$.
Since codewords are zero eigenvectors, a rank-deficient $H$ implies a large code space, with $k = n - \rank(H)$.
A rank-deficient seed code gives rise to superselection sectors in the resulting quantum code. 
A superselection sector is an equivalence class of syndromes under local errors; as the number of sectors grows, a diminishing number of excitation configurations may be dynamically connected by \emph{any} physical process~\footnote{As in topological order, superselection sectors identify dynamically disconnected excitation patterns; the difference is that rather than supporting finitely many, labeled by elements of the anyon theory, in an HGP code based on a rank-deficient seed code, the number of superselection sectors depends on the system size.}.
This constitutes a strong mechanism for immobility, and is central to both local and nonlocal fracton models~\cite{Kim_PRL2016}.

Another source of immobility is \emph{confinement}, a property related to glassiness which is studied in both classical and quantum codes~\cite{Bombin_PRX2015, Quintavalle_PRX2021}.
A typical classical LDPC code is linearly confining with high probability, as nonlocal checks lead individual errors to produce roughly $\kappa_b$ excitations.
Except in special cases~\cite{Yoshida_PRB2013}, if both seed codes are confining, then so is the resulting quantum code. 

An example of a classical seed code satisfying the prior conditions is the nonlocal Ising model; however the HGP of two such codes is not a fracton model.
This is a consequence of supporting only excitations which are extended objects~\cite{Bombin_PRX2015}, however, and is special within the space of nonlocal codes.
In order to exclude these cases, we impose an \emph{isolation} condition requiring an individual excitation to be point-like; that is, it can be separated from other excitations with high probability. 

Isolability is a natural feature of nonlocal codes, as a simple probabilistic argument demonstrates~\cite{Supplemental}.
This can be seen as a consequence of the unlikelihood of an ``Ising subgraph'': a set of bits on the Tanner graph whose neighboring checks are all two-valent.
Such an Ising subgraph contains a set of local symmetries constraining excitations to be extended domain walls, and thus not isolable~\footnote{In particular, this refers to a set of classical meta-checks, one for each independent cycle in the subgraph.}. 
Crucially, typical LDPC codes with $\kappa_c > 2$ are isolable, as the likelihood of large Ising subgraphs is exponentially suppressed.
As was the case with confinement, a HGP code inherits isolability only if this condition is satisfied in both seed codes.

\emph{Nonlocal Type-I fractons in HGP codes.---}We consider HGP codes, in which one classical seed is a cyclic repetition code and the other is either: case (i), a \textit{Laplacian code}; or case (ii), a typical LDPC code (see~\cite{Supplemental} for details of each).
This is motivated by a recently proposed Type-I fracton model called the ``anisotropic $\Z_2$ Laplacian model''~\cite{Gorantla_PRB2023,Ebisu_Scipost2023}.
While originally derived from a series of results on dipole-conserving tensor gauge theories~\cite{Seiberg_Exotic2021,Gorlanta_Modified2021,Gorantla_PRB2022}, this model is precisely described by case (i) above [Fig.~\ref{fig:venn}(c)]~\cite{Supplemental}.

\begin{figure}
\centering
\includegraphics[width=\columnwidth]{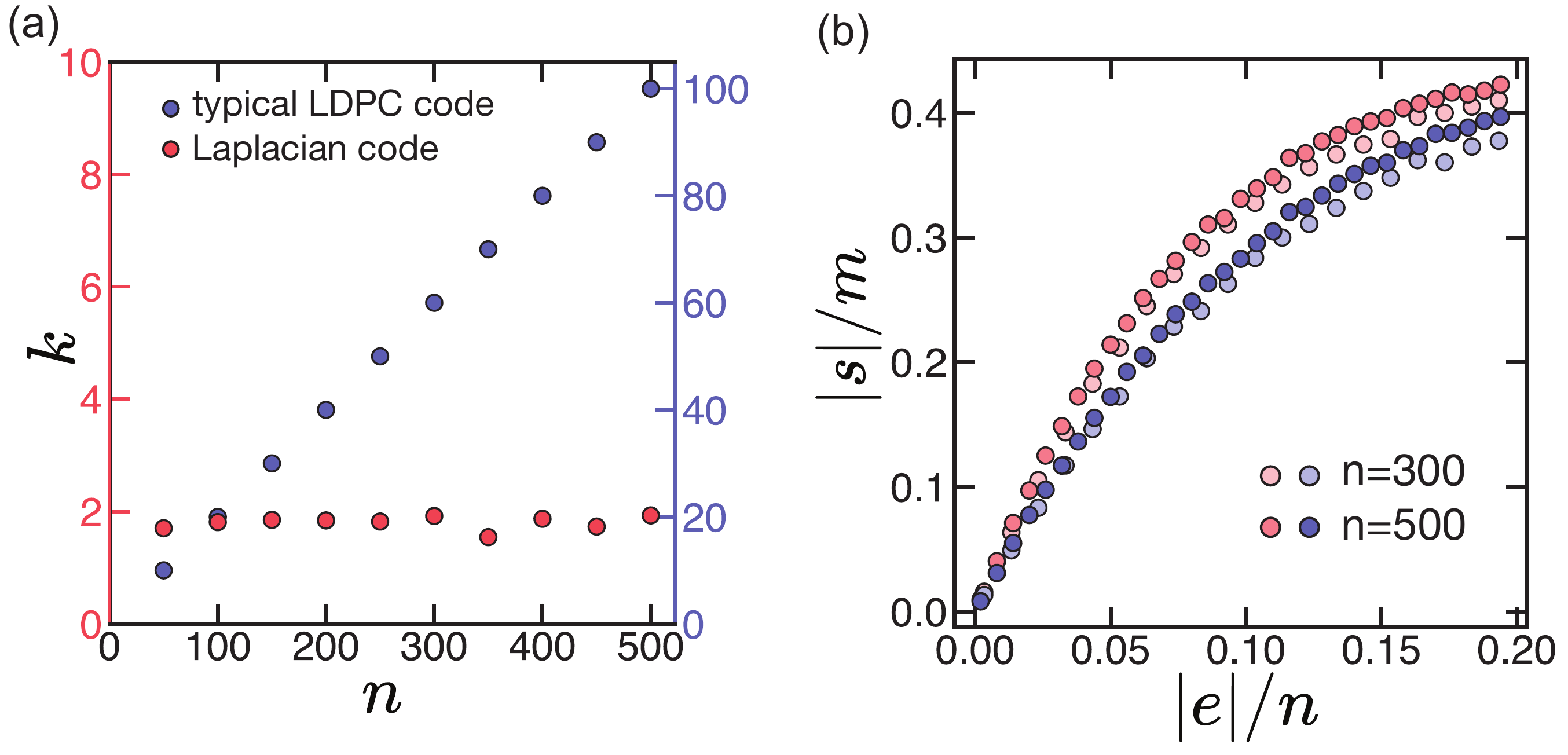}
\caption{\label{fig:laplacian} (a) Rank deficiency is exhibited for typical classical LDPC codes, from the linear scaling of $k$ with $n$; by contrast, the number of logical bits in a Laplacian code is constant.
(b) Confinement data is shown for both typical LDPC codes and Laplacian codes for $n=300$ and $500$.
Pictured is the minimum syndrome weight of randomly sampled errors.
Both constructions display evidence of confinement, as expected for nonlocal codes.
}
\end{figure}

The parity check matrix of the classical Laplacian code used in case (i) is given by the reduction mod 2 of the Laplacian operator of a graph~\cite{Chung_1997}.
To study its rank deficiency, we find the number of encoded logical bits $k$ as a function of system size $n$.
For each system size, we average over $\sim 10^3$ random graphs, showing the results in Fig.~\ref{fig:laplacian}(a).
It is immediately evident that the Laplacian code on a general sparse graph does not exhibit evidence of rank deficiency.
As a consequence, the resulting quantum code does not support fracton order.

This is in apparent contrast with results on the square lattice~\cite{Gorantla_PRB2023}, but can be understood from the perspective of the seed code.
Indeed, typical LDPC codes used in case (ii) are chosen with an imbalance between bits and checks leading to $k = O(n)$ rank deficiency, as shown in Fig.~\ref{fig:laplacian}(a).
Because the parity check matrix of the Laplacian code is square, it is very likely to be nearly full rank.
The square lattice is thus a special case, and exact results can in fact illuminate how translation invariance leads to rank deficiency in the Laplacian code~\cite{Supplemental}.

In contrast, we find that confinement is a feature of both typical LDPC and Laplacian codes.
We compute the minimum density $|s|/m$ of excitations induced by $\sim 10^3$ randomly sampled error configurations at each sparsity $|e|/n$.
The monotonic growth illustrated in Fig.~\ref{fig:laplacian}(b) suggests that both typical LDPC and Laplacian codes are confining.
Finally, we observe that both typical LDPC and Laplacian codes are isolable due to the suppression of Ising subgraphs by taking $\kappa > 2$~\cite{Supplemental}.

Thus, we conclude that typical LDPC codes are rank-deficient, confining, and isolable, whereas Laplacian codes on general graphs are merely confining and isolable.
Because the cyclic repetition code is also isolable, our case (ii) HGP code indeed realizes Type-I fractons.
However, rather than fractons, the case (i) HGP construction instead hosts partially confined point-like excitations in a phase akin to nonlocal topological order.

\emph{Local fracton models from aperiodic codes}.---
Though the stabilizer models discussed so far are only $\kappa$-local, it is natural to ask whether geometrically local HGP codes realize local fracton order.
Whereas one well-known route to locality is to define a Tanner graph on a negatively curved manifold, which can embed expanders~\cite{Delfosse_ISIT2013}, we instead take a Euclidean approach by placing both bits and checks on vertices of a two-dimensional aperiodic tiling.

As a specific example, we consider the pinwheel tiling, which has no exact spatial symmetries~\cite{Radin_Space1992,Radin_1994Pinwheel,Radin_Aperiodic1997,Frettloh_Substitution2008}.
This tiling is specified by a substitution rule [Fig.~\ref{fig:pinwheel_code}(a)] generating a family of graphs $\{G_N\}$, $N=1,2,\ldots$, whose volume scales as $n \sim e^N$ [see Fig.~\ref{fig:pinwheel_code}(b,c)].
We define parity check matrices $H_N$ based on the graph Laplacian $L_N$ at generation $N$~\cite{Chung_1997}.
We first compute $\tilde H_N = (L_N -\I)\mod 2$, where $\I$ is the $n\times n$ identity matrix.
Then $H_N$ is obtained via a ``boundary depletion'' of checks, truncating the column space of $\tilde H_N$ by removing an evenly spaced fraction $1/p$ of checks on the boundary vertices of $G_N$.
We dub this the \emph{pinwheel code}, within a more general class of \emph{aperiodic codes}.

At issue are thus the rank deficiency and confinement of the pinwheel code; isolability is guaranteed, as the connectivity of bits and checks ranges between three and nine for all generations.
As shown in Fig.~\ref{fig:pinwheel_code}(d), this family exhibits $k \approx \sqrt{n}/p$ and is thus rank deficient, with the parameter $p$ tuning the number of logical bits~\footnote{We observe that for small values of $p$ short logical operators attached to the boundary can appear, thus we take large enough values to suppress these.}.
Interestingly, although the depletion of checks relative to bits is a boundary effect, this does not imply that logical bits are localized on the boundary~\cite{Supplemental}.
Indeed, we find that this is not the case, measuring a code distance that scales linearly with $n$, as shown in Fig.~\ref{fig:pinwheel_code}(d).
The pinwheel code introduced here thus saturates the classical BPT bound $k\sqrt d = O(n)$~\cite{Bravyi_PRL2010}.

In contrast to typical LDPC codes, a $D$-dimensional local code cannot exhibit linear confinement.
An upper bound is realized by the Ising model, wherein domain walls have polynomial energy cost with exponent $\frac{D-1}{D}$; more generally, in any local code, a codeword can be truncated within a $D$-ball at the cost of only a $(D-1)$-dimensional boundary syndrome.
Such scaling $|s| \sim |e|^{1/2}$ is exhibited in the limit of large error weights in Fig.~\ref{fig:pinwheel_code}(e).
Though sampling is not guaranteed to capture the tails of the distribution, it is sufficiently thorough to distinguish the present cases from the Laplacian code on the square lattice, which, though rank deficient, is not confining~\cite{Supplemental}.
We thus conjecture that the pinwheel code is confining, and saturates the upper bound.
 
Having demonstrated rank deficiency, confinement, and isolability, we now consider the HGP of pinwheel codes as local fracton models. 
First, a pinwheel code and a cyclic repetition code yield a local Type-I fracton model in three dimensions, with translation symmetry perpendicular to the tiling only. 
Second, two instances of pinwheel codes yield a local Type-II fracton model in four dimensions, with parameters $k_q\sim n_q^{1/4}$, $d_q \sim \sqrt{n_q}$~\footnote{While we have studied a particular rectangular patch under the substitution rules, more general regions are possible, and due to the lack of spatial symmetry non-identical codes will not have large coincident subgraphs.}.
We therefore largely capture the physics associated with the typical LDPC code [case (ii)] without appealing to nonlocal interactions or disorder.
We conjecture that, similar to the four-dimensional toric code, the Type-II pinwheel stabilizer model is confining without disorder, but, crucially, achieves this for point-like rather than extended excitations.

\emph{Discussion and conclusion}.---
From the perspective of the present work, the study of quantum glassiness leading to fracton order can be understood as an exchange of confinement for superselection, each of which provides a complementary approach to constraining mobility~\cite{Kim_PRL2016}.
Here, we consider neither to be fundamentally classical or quantum, and instead characterize both as aspects of quantum product codes.

Looking forward, our work opens the door to a number of intriguing directions. 
First, more sophisticated product codes include the lifted and balanced products, and achieve a dimensional reduction by making use of lattice symmetries of the seed codes~\cite{Hastings_arxiv2021,Panteleev_ITI2022,Breuckmann_TIT2021}.
It is not immediately evident how to extend such an operation to aperiodic codes; however, this could in principle reduce a code similar to our local Type-II model to three dimensions, which could be compared to other recent constructions~\cite{Williamson_Layer2023}.
Second, it would be interesting to characterize the conditions on aperiodic tilings that lead to local Type-II fractons, and whether such tilings all flow to the same fixed point under entanglement renormalization~\cite{Vidal_ER2007,Haah_PRB2014,Dua_PRR2020,Dua_PRB2021}.
One possible avenue is an understanding in terms of condensation of certain topological lattice defects~\cite{Aitchison_Strings2023}.
Finally, it would be useful to quantitatively investigate the spin-glass contribution to immobility and the energy barrier in product code fracton models.

\emph{Acknowledgements}.---We gratefully acknowledge the insight of and discussions with  Hong-Ye Hu, Zhu-Xi Luo, Shu-Heng Shao, Thomas Schuster, and Carolyn Zhang.
We are particularly indebted to Helia Kamal for discussions leading to the origination of this project.
This work was supported in part by multiple grants from the  U.S. Department of Energy, including  the Accelerated Research in Quantum Computing (ARQC) program, the Quantum Information
Science Enabled Discovery (QuantISED) for High Energy
Physics (KA2401032) and through the GeoFlow Grant
No.~de-sc0019380. 
B.~R.~acknowledges support from the Harvard Quantum Initiative Postdoctoral Fellowship in Science Engineering.
N.~Y.~Y.~acknowledges support from a Simons Investigator award. 

\bibliography{refs}

\begin{thebibliography}{69}%
\makeatletter
\providecommand \@ifxundefined [1]{%
 \@ifx{#1\undefined}
}%
\providecommand \@ifnum [1]{%
 \ifnum #1\expandafter \@firstoftwo
 \else \expandafter \@secondoftwo
 \fi
}%
\providecommand \@ifx [1]{%
 \ifx #1\expandafter \@firstoftwo
 \else \expandafter \@secondoftwo
 \fi
}%
\providecommand \natexlab [1]{#1}%
\providecommand \enquote  [1]{``#1''}%
\providecommand \bibnamefont  [1]{#1}%
\providecommand \bibfnamefont [1]{#1}%
\providecommand \citenamefont [1]{#1}%
\providecommand \href@noop [0]{\@secondoftwo}%
\providecommand \href [0]{\begingroup \@sanitize@url \@href}%
\providecommand \@href[1]{\@@startlink{#1}\@@href}%
\providecommand \@@href[1]{\endgroup#1\@@endlink}%
\providecommand \@sanitize@url [0]{\catcode `\\12\catcode `\$12\catcode `\&12\catcode `\#12\catcode `\^12\catcode `\_12\catcode `\%12\relax}%
\providecommand \@@startlink[1]{}%
\providecommand \@@endlink[0]{}%
\providecommand \url  [0]{\begingroup\@sanitize@url \@url }%
\providecommand \@url [1]{\endgroup\@href {#1}{\urlprefix }}%
\providecommand \urlprefix  [0]{URL }%
\providecommand \Eprint [0]{\href }%
\providecommand \doibase [0]{https://doi.org/}%
\providecommand \selectlanguage [0]{\@gobble}%
\providecommand \bibinfo  [0]{\@secondoftwo}%
\providecommand \bibfield  [0]{\@secondoftwo}%
\providecommand \translation [1]{[#1]}%
\providecommand \BibitemOpen [0]{}%
\providecommand \bibitemStop [0]{}%
\providecommand \bibitemNoStop [0]{.\EOS\space}%
\providecommand \EOS [0]{\spacefactor3000\relax}%
\providecommand \BibitemShut  [1]{\csname bibitem#1\endcsname}%
\let\auto@bib@innerbib\@empty
\bibitem [{\citenamefont {Chamon}(2005)}]{Chamon_PRL2005}%
  \BibitemOpen
  \bibfield  {author} {\bibinfo {author} {\bibfnamefont {C.}~\bibnamefont {Chamon}},\ }\bibfield  {title} {\bibinfo {title} {Quantum glassiness in strongly correlated clean systems: An example of topological overprotection},\ }\href {https://doi.org/10.1103/PhysRevLett.94.040402} {\bibfield  {journal} {\bibinfo  {journal} {Phys. Rev. Lett.}\ }\textbf {\bibinfo {volume} {94}},\ \bibinfo {pages} {040402} (\bibinfo {year} {2005})}\BibitemShut {NoStop}%
\bibitem [{\citenamefont {Haah}(2011)}]{Haah_PRA2011}%
  \BibitemOpen
  \bibfield  {author} {\bibinfo {author} {\bibfnamefont {J.}~\bibnamefont {Haah}},\ }\bibfield  {title} {\bibinfo {title} {Local stabilizer codes in three dimensions without string logical operators},\ }\href {https://doi.org/10.1103/PhysRevA.83.042330} {\bibfield  {journal} {\bibinfo  {journal} {Phys. Rev. A}\ }\textbf {\bibinfo {volume} {83}},\ \bibinfo {pages} {042330} (\bibinfo {year} {2011})}\BibitemShut {NoStop}%
\bibitem [{\citenamefont {Yoshida}(2013)}]{Yoshida_PRB2013}%
  \BibitemOpen
  \bibfield  {author} {\bibinfo {author} {\bibfnamefont {B.}~\bibnamefont {Yoshida}},\ }\bibfield  {title} {\bibinfo {title} {Exotic topological order in fractal spin liquids},\ }\href {https://doi.org/10.1103/PhysRevB.88.125122} {\bibfield  {journal} {\bibinfo  {journal} {Phys. Rev. B}\ }\textbf {\bibinfo {volume} {88}},\ \bibinfo {pages} {125122} (\bibinfo {year} {2013})}\BibitemShut {NoStop}%
\bibitem [{\citenamefont {Vijay}\ \emph {et~al.}(2015)\citenamefont {Vijay}, \citenamefont {Haah},\ and\ \citenamefont {Fu}}]{Vijay_PRB2015}%
  \BibitemOpen
  \bibfield  {author} {\bibinfo {author} {\bibfnamefont {S.}~\bibnamefont {Vijay}}, \bibinfo {author} {\bibfnamefont {J.}~\bibnamefont {Haah}},\ and\ \bibinfo {author} {\bibfnamefont {L.}~\bibnamefont {Fu}},\ }\bibfield  {title} {\bibinfo {title} {A new kind of topological quantum order: A dimensional hierarchy of quasiparticles built from stationary excitations},\ }\href {https://doi.org/10.1103/PhysRevB.92.235136} {\bibfield  {journal} {\bibinfo  {journal} {Phys. Rev. B}\ }\textbf {\bibinfo {volume} {92}},\ \bibinfo {pages} {235136} (\bibinfo {year} {2015})}\BibitemShut {NoStop}%
\bibitem [{\citenamefont {Vijay}\ \emph {et~al.}(2016)\citenamefont {Vijay}, \citenamefont {Haah},\ and\ \citenamefont {Fu}}]{Vijay_PRB2016}%
  \BibitemOpen
  \bibfield  {author} {\bibinfo {author} {\bibfnamefont {S.}~\bibnamefont {Vijay}}, \bibinfo {author} {\bibfnamefont {J.}~\bibnamefont {Haah}},\ and\ \bibinfo {author} {\bibfnamefont {L.}~\bibnamefont {Fu}},\ }\bibfield  {title} {\bibinfo {title} {Fracton topological order, generalized lattice gauge theory, and duality},\ }\href {https://doi.org/10.1103/PhysRevB.94.235157} {\bibfield  {journal} {\bibinfo  {journal} {Phys. Rev. B}\ }\textbf {\bibinfo {volume} {94}},\ \bibinfo {pages} {235157} (\bibinfo {year} {2016})}\BibitemShut {NoStop}%
\bibitem [{\citenamefont {Slagle}\ and\ \citenamefont {Kim}(2017)}]{Slagle_PRB2017}%
  \BibitemOpen
  \bibfield  {author} {\bibinfo {author} {\bibfnamefont {K.}~\bibnamefont {Slagle}}\ and\ \bibinfo {author} {\bibfnamefont {Y.~B.}\ \bibnamefont {Kim}},\ }\bibfield  {title} {\bibinfo {title} {Quantum field theory of {$X$}-cube fracton topological order and robust degeneracy from geometry},\ }\href {https://doi.org/10.1103/PhysRevB.96.195139} {\bibfield  {journal} {\bibinfo  {journal} {Phys. Rev. B}\ }\textbf {\bibinfo {volume} {96}},\ \bibinfo {pages} {195139} (\bibinfo {year} {2017})}\BibitemShut {NoStop}%
\bibitem [{\citenamefont {Pretko}(2018)}]{Pretko_PRB2018}%
  \BibitemOpen
  \bibfield  {author} {\bibinfo {author} {\bibfnamefont {M.}~\bibnamefont {Pretko}},\ }\bibfield  {title} {\bibinfo {title} {The fracton gauge principle},\ }\href {https://doi.org/10.1103/PhysRevB.98.115134} {\bibfield  {journal} {\bibinfo  {journal} {Phys. Rev. B}\ }\textbf {\bibinfo {volume} {98}},\ \bibinfo {pages} {115134} (\bibinfo {year} {2018})}\BibitemShut {NoStop}%
\bibitem [{\citenamefont {Gromov}(2019)}]{Gromov_PRX2019}%
  \BibitemOpen
  \bibfield  {author} {\bibinfo {author} {\bibfnamefont {A.}~\bibnamefont {Gromov}},\ }\bibfield  {title} {\bibinfo {title} {Towards classification of fracton phases: The multipole algebra},\ }\href {https://doi.org/10.1103/PhysRevX.9.031035} {\bibfield  {journal} {\bibinfo  {journal} {Phys. Rev. X}\ }\textbf {\bibinfo {volume} {9}},\ \bibinfo {pages} {031035} (\bibinfo {year} {2019})}\BibitemShut {NoStop}%
\bibitem [{\citenamefont {Shirley}\ \emph {et~al.}(2019{\natexlab{a}})\citenamefont {Shirley}, \citenamefont {Slagle},\ and\ \citenamefont {Chen}}]{Shirley_Foliated2019}%
  \BibitemOpen
  \bibfield  {author} {\bibinfo {author} {\bibfnamefont {W.}~\bibnamefont {Shirley}}, \bibinfo {author} {\bibfnamefont {K.}~\bibnamefont {Slagle}},\ and\ \bibinfo {author} {\bibfnamefont {X.}~\bibnamefont {Chen}},\ }\bibfield  {title} {\bibinfo {title} {{Foliated fracton order from gauging subsystem symmetries}},\ }\href {https://doi.org/10.21468/SciPostPhys.6.4.041} {\bibfield  {journal} {\bibinfo  {journal} {SciPost Phys.}\ }\textbf {\bibinfo {volume} {6}},\ \bibinfo {pages} {041} (\bibinfo {year} {2019}{\natexlab{a}})}\BibitemShut {NoStop}%
\bibitem [{\citenamefont {Dua}\ \emph {et~al.}(2019)\citenamefont {Dua}, \citenamefont {Kim}, \citenamefont {Cheng},\ and\ \citenamefont {Williamson}}]{Dua_Sorting2019}%
  \BibitemOpen
  \bibfield  {author} {\bibinfo {author} {\bibfnamefont {A.}~\bibnamefont {Dua}}, \bibinfo {author} {\bibfnamefont {I.~H.}\ \bibnamefont {Kim}}, \bibinfo {author} {\bibfnamefont {M.}~\bibnamefont {Cheng}},\ and\ \bibinfo {author} {\bibfnamefont {D.~J.}\ \bibnamefont {Williamson}},\ }\bibfield  {title} {\bibinfo {title} {Sorting topological stabilizer models in three dimensions},\ }\href {https://doi.org/10.1103/PhysRevB.100.155137} {\bibfield  {journal} {\bibinfo  {journal} {Phys. Rev. B}\ }\textbf {\bibinfo {volume} {100}},\ \bibinfo {pages} {155137} (\bibinfo {year} {2019})}\BibitemShut {NoStop}%
\bibitem [{\citenamefont {Nandkishore}\ and\ \citenamefont {Hermele}(2019)}]{Nandkishore_ARCMP2019}%
  \BibitemOpen
  \bibfield  {author} {\bibinfo {author} {\bibfnamefont {R.~M.}\ \bibnamefont {Nandkishore}}\ and\ \bibinfo {author} {\bibfnamefont {M.}~\bibnamefont {Hermele}},\ }\bibfield  {title} {\bibinfo {title} {Fractons},\ }\href {https://doi.org/10.1146/annurev-conmatphys-031218-013604} {\bibfield  {journal} {\bibinfo  {journal} {Annual Review of Condensed Matter Physics}\ }\textbf {\bibinfo {volume} {10}},\ \bibinfo {pages} {295} (\bibinfo {year} {2019})},\ \Eprint {https://arxiv.org/abs/https://doi.org/10.1146/annurev-conmatphys-031218-013604} {https://doi.org/10.1146/annurev-conmatphys-031218-013604} \BibitemShut {NoStop}%
\bibitem [{\citenamefont {Pretko}\ \emph {et~al.}(2020)\citenamefont {Pretko}, \citenamefont {Chen},\ and\ \citenamefont {You}}]{Pretko_IJMPA2020}%
  \BibitemOpen
  \bibfield  {author} {\bibinfo {author} {\bibfnamefont {M.}~\bibnamefont {Pretko}}, \bibinfo {author} {\bibfnamefont {X.}~\bibnamefont {Chen}},\ and\ \bibinfo {author} {\bibfnamefont {Y.}~\bibnamefont {You}},\ }\bibfield  {title} {\bibinfo {title} {Fracton phases of matter},\ }\href {https://doi.org/10.1142/S0217751X20300033} {\bibfield  {journal} {\bibinfo  {journal} {International Journal of Modern Physics A}\ }\textbf {\bibinfo {volume} {35}},\ \bibinfo {pages} {2030003} (\bibinfo {year} {2020})},\ \Eprint {https://arxiv.org/abs/https://doi.org/10.1142/S0217751X20300033} {https://doi.org/10.1142/S0217751X20300033} \BibitemShut {NoStop}%
\bibitem [{\citenamefont {Breuckmann}\ and\ \citenamefont {Eberhardt}(2021{\natexlab{a}})}]{Breuckmann_TIT2021}%
  \BibitemOpen
  \bibfield  {author} {\bibinfo {author} {\bibfnamefont {N.~P.}\ \bibnamefont {Breuckmann}}\ and\ \bibinfo {author} {\bibfnamefont {J.~N.}\ \bibnamefont {Eberhardt}},\ }\bibfield  {title} {\bibinfo {title} {Balanced product quantum codes},\ }\href {https://doi.org/10.1109/TIT.2021.3097347} {\bibfield  {journal} {\bibinfo  {journal} {IEEE Transactions on Information Theory}\ }\textbf {\bibinfo {volume} {67}},\ \bibinfo {pages} {6653} (\bibinfo {year} {2021}{\natexlab{a}})}\BibitemShut {NoStop}%
\bibitem [{\citenamefont {Panteleev}\ and\ \citenamefont {Kalachev}(2022{\natexlab{a}})}]{Panteleev_ACM2022}%
  \BibitemOpen
  \bibfield  {author} {\bibinfo {author} {\bibfnamefont {P.}~\bibnamefont {Panteleev}}\ and\ \bibinfo {author} {\bibfnamefont {G.}~\bibnamefont {Kalachev}},\ }\bibfield  {title} {\bibinfo {title} {Asymptotically good quantum and locally testable classical {LDPC} codes},\ }in\ \href {https://doi.org/10.1145/3519935.3520017} {\emph {\bibinfo {booktitle} {Proceedings of the 54th Annual ACM SIGACT Symposium on Theory of Computing}}},\ \bibinfo {series and number} {STOC 2022}\ (\bibinfo  {publisher} {Association for Computing Machinery},\ \bibinfo {address} {New York, NY, USA},\ \bibinfo {year} {2022})\ p.\ \bibinfo {pages} {375–388}\BibitemShut {NoStop}%
\bibitem [{\citenamefont {Breuckmann}\ and\ \citenamefont {Eberhardt}(2021{\natexlab{b}})}]{Breuckmann_PRX2021}%
  \BibitemOpen
  \bibfield  {author} {\bibinfo {author} {\bibfnamefont {N.~P.}\ \bibnamefont {Breuckmann}}\ and\ \bibinfo {author} {\bibfnamefont {J.~N.}\ \bibnamefont {Eberhardt}},\ }\bibfield  {title} {\bibinfo {title} {Quantum low-density parity-check codes},\ }\href {https://doi.org/10.1103/PRXQuantum.2.040101} {\bibfield  {journal} {\bibinfo  {journal} {PRX Quantum}\ }\textbf {\bibinfo {volume} {2}},\ \bibinfo {pages} {040101} (\bibinfo {year} {2021}{\natexlab{b}})}\BibitemShut {NoStop}%
\bibitem [{Sup()}]{Supplemental}%
  \BibitemOpen
  \href@noop {} {\bibinfo {title} {Supplemental material}}\BibitemShut {NoStop}%
\bibitem [{\citenamefont {Gorantla}\ \emph {et~al.}(2023)\citenamefont {Gorantla}, \citenamefont {Lam}, \citenamefont {Seiberg},\ and\ \citenamefont {Shao}}]{Gorantla_PRB2023}%
  \BibitemOpen
  \bibfield  {author} {\bibinfo {author} {\bibfnamefont {P.}~\bibnamefont {Gorantla}}, \bibinfo {author} {\bibfnamefont {H.~T.}\ \bibnamefont {Lam}}, \bibinfo {author} {\bibfnamefont {N.}~\bibnamefont {Seiberg}},\ and\ \bibinfo {author} {\bibfnamefont {S.-H.}\ \bibnamefont {Shao}},\ }\bibfield  {title} {\bibinfo {title} {Gapped lineon and fracton models on graphs},\ }\href {https://doi.org/10.1103/PhysRevB.107.125121} {\bibfield  {journal} {\bibinfo  {journal} {Phys. Rev. B}\ }\textbf {\bibinfo {volume} {107}},\ \bibinfo {pages} {125121} (\bibinfo {year} {2023})}\BibitemShut {NoStop}%
\bibitem [{\citenamefont {Ebisu}\ and\ \citenamefont {Han}(2023)}]{Ebisu_Scipost2023}%
  \BibitemOpen
  \bibfield  {author} {\bibinfo {author} {\bibfnamefont {H.}~\bibnamefont {Ebisu}}\ and\ \bibinfo {author} {\bibfnamefont {B.}~\bibnamefont {Han}},\ }\bibfield  {title} {\bibinfo {title} {{Anisotropic higher rank $\mathbb{Z}_N$ topological phases on graphs}},\ }\href {https://doi.org/10.21468/SciPostPhys.14.5.106} {\bibfield  {journal} {\bibinfo  {journal} {SciPost Phys.}\ }\textbf {\bibinfo {volume} {14}},\ \bibinfo {pages} {106} (\bibinfo {year} {2023})}\BibitemShut {NoStop}%
\bibitem [{\citenamefont {Hastings}\ \emph {et~al.}(2020)\citenamefont {Hastings}, \citenamefont {Haah},\ and\ \citenamefont {O'Donnell}}]{Hastings_arxiv2021}%
  \BibitemOpen
  \bibfield  {author} {\bibinfo {author} {\bibfnamefont {M.~B.}\ \bibnamefont {Hastings}}, \bibinfo {author} {\bibfnamefont {J.}~\bibnamefont {Haah}},\ and\ \bibinfo {author} {\bibfnamefont {R.}~\bibnamefont {O'Donnell}},\ }\href@noop {} {\bibinfo {title} {Fiber bundle codes: Breaking the $n^{1/2} \operatorname{polylog}(n)$ barrier for quantum ldpc codes}} (\bibinfo {year} {2020}),\ \Eprint {https://arxiv.org/abs/2009.03921} {arXiv:2009.03921 [quant-ph]} \BibitemShut {NoStop}%
\bibitem [{\citenamefont {Panteleev}\ and\ \citenamefont {Kalachev}(2022{\natexlab{b}})}]{Panteleev_ITI2022}%
  \BibitemOpen
  \bibfield  {author} {\bibinfo {author} {\bibfnamefont {P.}~\bibnamefont {Panteleev}}\ and\ \bibinfo {author} {\bibfnamefont {G.}~\bibnamefont {Kalachev}},\ }\bibfield  {title} {\bibinfo {title} {Quantum {LDPC} codes with almost linear minimum distance},\ }\href {https://doi.org/10.1109/TIT.2021.3119384} {\bibfield  {journal} {\bibinfo  {journal} {IEEE Transactions on Information Theory}\ }\textbf {\bibinfo {volume} {68}},\ \bibinfo {pages} {213} (\bibinfo {year} {2022}{\natexlab{b}})}\BibitemShut {NoStop}%
\bibitem [{\citenamefont {Tillich}\ and\ \citenamefont {Zémor}(2014)}]{Tillich_TIT2014}%
  \BibitemOpen
  \bibfield  {author} {\bibinfo {author} {\bibfnamefont {J.-P.}\ \bibnamefont {Tillich}}\ and\ \bibinfo {author} {\bibfnamefont {G.}~\bibnamefont {Zémor}},\ }\bibfield  {title} {\bibinfo {title} {Quantum {LDPC} codes with positive rate and minimum distance proportional to the square root of the blocklength},\ }\href {https://doi.org/10.1109/TIT.2013.2292061} {\bibfield  {journal} {\bibinfo  {journal} {IEEE Transactions on Information Theory}\ }\textbf {\bibinfo {volume} {60}},\ \bibinfo {pages} {1193} (\bibinfo {year} {2014})}\BibitemShut {NoStop}%
\bibitem [{\citenamefont {Haah}(2013)}]{Haah_Commuting2013}%
  \BibitemOpen
  \bibfield  {author} {\bibinfo {author} {\bibfnamefont {J.}~\bibnamefont {Haah}},\ }\bibfield  {title} {\bibinfo {title} {Commuting {P}auli {H}amiltonians as maps between free modules},\ }\href {https://doi.org/10.1007/s00220-013-1810-2} {\bibfield  {journal} {\bibinfo  {journal} {Communications in Mathematical Physics}\ }\textbf {\bibinfo {volume} {324}},\ \bibinfo {pages} {351} (\bibinfo {year} {2013})}\BibitemShut {NoStop}%
\bibitem [{\citenamefont {Haah}(2016)}]{Haah_Algebraic2016}%
  \BibitemOpen
  \bibfield  {author} {\bibinfo {author} {\bibfnamefont {J.}~\bibnamefont {Haah}},\ }\bibfield  {title} {\bibinfo {title} {Algebraic methods for quantum codes on lattices},\ }\href {https://doi.org/10.15446/recolma.v50n2.62214} {\bibfield  {journal} {\bibinfo  {journal} {Revista colombiana de matematicas}\ }\textbf {\bibinfo {volume} {50}},\ \bibinfo {pages} {299} (\bibinfo {year} {2016})}\BibitemShut {NoStop}%
\bibitem [{\citenamefont {Bomb\'{\i}n}(2015)}]{Bombin_PRX2015}%
  \BibitemOpen
  \bibfield  {author} {\bibinfo {author} {\bibfnamefont {H.}~\bibnamefont {Bomb\'{\i}n}},\ }\bibfield  {title} {\bibinfo {title} {Single-shot fault-tolerant quantum error correction},\ }\href {https://doi.org/10.1103/PhysRevX.5.031043} {\bibfield  {journal} {\bibinfo  {journal} {Phys. Rev. X}\ }\textbf {\bibinfo {volume} {5}},\ \bibinfo {pages} {031043} (\bibinfo {year} {2015})}\BibitemShut {NoStop}%
\bibitem [{\citenamefont {Quintavalle}\ \emph {et~al.}(2021)\citenamefont {Quintavalle}, \citenamefont {Vasmer}, \citenamefont {Roffe},\ and\ \citenamefont {Campbell}}]{Quintavalle_PRX2021}%
  \BibitemOpen
  \bibfield  {author} {\bibinfo {author} {\bibfnamefont {A.~O.}\ \bibnamefont {Quintavalle}}, \bibinfo {author} {\bibfnamefont {M.}~\bibnamefont {Vasmer}}, \bibinfo {author} {\bibfnamefont {J.}~\bibnamefont {Roffe}},\ and\ \bibinfo {author} {\bibfnamefont {E.~T.}\ \bibnamefont {Campbell}},\ }\bibfield  {title} {\bibinfo {title} {Single-shot error correction of three-dimensional homological product codes},\ }\href {https://doi.org/10.1103/PRXQuantum.2.020340} {\bibfield  {journal} {\bibinfo  {journal} {PRX Quantum}\ }\textbf {\bibinfo {volume} {2}},\ \bibinfo {pages} {020340} (\bibinfo {year} {2021})}\BibitemShut {NoStop}%
\bibitem [{\citenamefont {MacKay}(2003)}]{MacKay_2003}%
  \BibitemOpen
  \bibfield  {author} {\bibinfo {author} {\bibfnamefont {D.~J.~C.}\ \bibnamefont {MacKay}},\ }\href@noop {} {\emph {\bibinfo {title} {Information Theory, Inference, and Learning Algorithms}}}\ (\bibinfo  {publisher} {Copyright Cambridge University Press},\ \bibinfo {year} {2003})\BibitemShut {NoStop}%
\bibitem [{\citenamefont {Calderbank}\ and\ \citenamefont {Shor}(1996)}]{Calderbank_PRA1996}%
  \BibitemOpen
  \bibfield  {author} {\bibinfo {author} {\bibfnamefont {A.~R.}\ \bibnamefont {Calderbank}}\ and\ \bibinfo {author} {\bibfnamefont {P.~W.}\ \bibnamefont {Shor}},\ }\bibfield  {title} {\bibinfo {title} {Good quantum error-correcting codes exist},\ }\href {https://doi.org/10.1103/PhysRevA.54.1098} {\bibfield  {journal} {\bibinfo  {journal} {Phys. Rev. A}\ }\textbf {\bibinfo {volume} {54}},\ \bibinfo {pages} {1098} (\bibinfo {year} {1996})}\BibitemShut {NoStop}%
\bibitem [{\citenamefont {Steane}(1996)}]{Steane_PRL1996}%
  \BibitemOpen
  \bibfield  {author} {\bibinfo {author} {\bibfnamefont {A.~M.}\ \bibnamefont {Steane}},\ }\bibfield  {title} {\bibinfo {title} {Error correcting codes in quantum theory},\ }\href {https://doi.org/10.1103/PhysRevLett.77.793} {\bibfield  {journal} {\bibinfo  {journal} {Phys. Rev. Lett.}\ }\textbf {\bibinfo {volume} {77}},\ \bibinfo {pages} {793} (\bibinfo {year} {1996})}\BibitemShut {NoStop}%
\bibitem [{\citenamefont {Tanner}(1981)}]{Tanner_IEEE1981}%
  \BibitemOpen
  \bibfield  {author} {\bibinfo {author} {\bibfnamefont {R.}~\bibnamefont {Tanner}},\ }\bibfield  {title} {\bibinfo {title} {A recursive approach to low complexity codes},\ }\href {https://doi.org/10.1109/TIT.1981.1056404} {\bibfield  {journal} {\bibinfo  {journal} {IEEE Transactions on Information Theory}\ }\textbf {\bibinfo {volume} {27}},\ \bibinfo {pages} {533} (\bibinfo {year} {1981})}\BibitemShut {NoStop}%
\bibitem [{Note1()}]{Note1}%
  \BibitemOpen
  \bibinfo {note} {Note that this scaling is distinct from that of local fractons, which merely follows a polynomial envelope.}\BibitemShut {Stop}%
\bibitem [{Note2()}]{Note2}%
  \BibitemOpen
  \bibinfo {note} {While finite-dimensional local fracton models with immobile loop excitations are known~\cite {LiYe20}, such a case is not expected for the nonlocal models we consider.}\BibitemShut {Stop}%
\bibitem [{Note3()}]{Note3}%
  \BibitemOpen
  \bibinfo {note} {The definition of confinement here is weaker than its usage elsewhere~\cite {Bombin_PRX2015}, as the energy cost is allowed to scale as any increasing function of the error weight, rather than strictly linearly.}\BibitemShut {Stop}%
\bibitem [{Note4()}]{Note4}%
  \BibitemOpen
  \bibinfo {note} {As in topological order, superselection sectors identify dynamically disconnected excitation patterns; the difference is that rather than supporting finitely many, labeled by elements of the anyon theory, in an HGP code based on a rank-deficient seed code, the number of superselection sectors depends on the system size.}\BibitemShut {Stop}%
\bibitem [{\citenamefont {Kim}\ and\ \citenamefont {Haah}(2016)}]{Kim_PRL2016}%
  \BibitemOpen
  \bibfield  {author} {\bibinfo {author} {\bibfnamefont {I.~H.}\ \bibnamefont {Kim}}\ and\ \bibinfo {author} {\bibfnamefont {J.}~\bibnamefont {Haah}},\ }\bibfield  {title} {\bibinfo {title} {Localization from superselection rules in translationally invariant systems},\ }\href {https://doi.org/10.1103/PhysRevLett.116.027202} {\bibfield  {journal} {\bibinfo  {journal} {Phys. Rev. Lett.}\ }\textbf {\bibinfo {volume} {116}},\ \bibinfo {pages} {027202} (\bibinfo {year} {2016})}\BibitemShut {NoStop}%
\bibitem [{Note5()}]{Note5}%
  \BibitemOpen
  \bibinfo {note} {In particular, this refers to a set of classical meta-checks, one for each independent cycle in the subgraph.}\BibitemShut {Stop}%
\bibitem [{\citenamefont {Seiberg}\ and\ \citenamefont {Shao}(2021)}]{Seiberg_Exotic2021}%
  \BibitemOpen
  \bibfield  {author} {\bibinfo {author} {\bibfnamefont {N.}~\bibnamefont {Seiberg}}\ and\ \bibinfo {author} {\bibfnamefont {S.-H.}\ \bibnamefont {Shao}},\ }\bibfield  {title} {\bibinfo {title} {{Exotic symmetries, duality, and fractons in 2+1-dimensional quantum field theory}},\ }\href {https://doi.org/10.21468/SciPostPhys.10.2.027} {\bibfield  {journal} {\bibinfo  {journal} {SciPost Phys.}\ }\textbf {\bibinfo {volume} {10}},\ \bibinfo {pages} {027} (\bibinfo {year} {2021})}\BibitemShut {NoStop}%
\bibitem [{\citenamefont {Gorantla}\ \emph {et~al.}(2021)\citenamefont {Gorantla}, \citenamefont {Lam}, \citenamefont {Seiberg},\ and\ \citenamefont {Shao}}]{Gorlanta_Modified2021}%
  \BibitemOpen
  \bibfield  {author} {\bibinfo {author} {\bibfnamefont {P.}~\bibnamefont {Gorantla}}, \bibinfo {author} {\bibfnamefont {H.~T.}\ \bibnamefont {Lam}}, \bibinfo {author} {\bibfnamefont {N.}~\bibnamefont {Seiberg}},\ and\ \bibinfo {author} {\bibfnamefont {S.-H.}\ \bibnamefont {Shao}},\ }\bibfield  {title} {\bibinfo {title} {{A modified Villain formulation of fractons and other exotic theories}},\ }\href {https://doi.org/10.1063/5.0060808} {\bibfield  {journal} {\bibinfo  {journal} {Journal of Mathematical Physics}\ }\textbf {\bibinfo {volume} {62}},\ \bibinfo {pages} {102301} (\bibinfo {year} {2021})},\ \Eprint {https://arxiv.org/abs/https://pubs.aip.org/aip/jmp/article-pdf/doi/10.1063/5.0060808/15934165/102301\_1\_online.pdf} {https://pubs.aip.org/aip/jmp/article-pdf/doi/10.1063/5.0060808/15934165/102301\_1\_online.pdf} \BibitemShut {NoStop}%
\bibitem [{\citenamefont {Gorantla}\ \emph {et~al.}(2022)\citenamefont {Gorantla}, \citenamefont {Lam},\ and\ \citenamefont {Shao}}]{Gorantla_PRB2022}%
  \BibitemOpen
  \bibfield  {author} {\bibinfo {author} {\bibfnamefont {P.}~\bibnamefont {Gorantla}}, \bibinfo {author} {\bibfnamefont {H.~T.}\ \bibnamefont {Lam}},\ and\ \bibinfo {author} {\bibfnamefont {S.-H.}\ \bibnamefont {Shao}},\ }\bibfield  {title} {\bibinfo {title} {Fractons on graphs and complexity},\ }\href {https://doi.org/10.1103/PhysRevB.106.195139} {\bibfield  {journal} {\bibinfo  {journal} {Phys. Rev. B}\ }\textbf {\bibinfo {volume} {106}},\ \bibinfo {pages} {195139} (\bibinfo {year} {2022})}\BibitemShut {NoStop}%
\bibitem [{\citenamefont {Chung}(1997)}]{Chung_1997}%
  \BibitemOpen
  \bibfield  {author} {\bibinfo {author} {\bibfnamefont {F.~R.}\ \bibnamefont {Chung}},\ }\href {https://doi.org/10.1090/cbms/092} {\emph {\bibinfo {title} {Spectral graph theory}}},\ Vol.~\bibinfo {volume} {92}\ (\bibinfo  {publisher} {American Mathematical Soc.},\ \bibinfo {year} {1997})\BibitemShut {NoStop}%
\bibitem [{\citenamefont {Delfosse}(2013)}]{Delfosse_ISIT2013}%
  \BibitemOpen
  \bibfield  {author} {\bibinfo {author} {\bibfnamefont {N.}~\bibnamefont {Delfosse}},\ }\bibfield  {title} {\bibinfo {title} {Tradeoffs for reliable quantum information storage in surface codes and color codes},\ }in\ \href {https://doi.org/10.1109/ISIT.2013.6620360} {\emph {\bibinfo {booktitle} {2013 IEEE International Symposium on Information Theory}}}\ (\bibinfo {year} {2013})\ pp.\ \bibinfo {pages} {917--921}\BibitemShut {NoStop}%
\bibitem [{\citenamefont {Radin}\ and\ \citenamefont {Wolff}(1992)}]{Radin_Space1992}%
  \BibitemOpen
  \bibfield  {author} {\bibinfo {author} {\bibfnamefont {C.}~\bibnamefont {Radin}}\ and\ \bibinfo {author} {\bibfnamefont {M.}~\bibnamefont {Wolff}},\ }\bibfield  {title} {\bibinfo {title} {Space tilings and local isomorphism},\ }\href {https://doi.org/10.1007/BF02414073} {\bibfield  {journal} {\bibinfo  {journal} {Geometriae Dedicata}\ }\textbf {\bibinfo {volume} {42}},\ \bibinfo {pages} {355} (\bibinfo {year} {1992})}\BibitemShut {NoStop}%
\bibitem [{\citenamefont {Radin}(1994)}]{Radin_1994Pinwheel}%
  \BibitemOpen
  \bibfield  {author} {\bibinfo {author} {\bibfnamefont {C.}~\bibnamefont {Radin}},\ }\bibfield  {title} {\bibinfo {title} {The pinwheel tilings of the plane},\ }\href {http://www.jstor.org/stable/2118575} {\bibfield  {journal} {\bibinfo  {journal} {Annals of Mathematics}\ }\textbf {\bibinfo {volume} {139}},\ \bibinfo {pages} {661} (\bibinfo {year} {1994})}\BibitemShut {NoStop}%
\bibitem [{\citenamefont {Radin}(1997)}]{Radin_Aperiodic1997}%
  \BibitemOpen
  \bibfield  {author} {\bibinfo {author} {\bibfnamefont {C.}~\bibnamefont {Radin}},\ }\bibfield  {title} {\bibinfo {title} {Aperiodic tilings, ergodic theory, and rotations},\ }\href {https://doi.org/10.1007/978-94-015-8784-6_19} {\bibfield  {journal} {\bibinfo  {journal} {NATO ASI Series C Mathematical and Physical Sciences-Advanced Study Institute}\ }\textbf {\bibinfo {volume} {489}},\ \bibinfo {pages} {499} (\bibinfo {year} {1997})}\BibitemShut {NoStop}%
\bibitem [{\citenamefont {Frettl\"{o}h}(2008)}]{Frettloh_Substitution2008}%
  \BibitemOpen
  \bibfield  {author} {\bibinfo {author} {\bibfnamefont {D.}~\bibnamefont {Frettl\"{o}h}},\ }\bibfield  {title} {\bibinfo {title} {Substitution tilings with statistical circular symmetry},\ }\href {https://doi.org/10.1016/j.ejc.2008.01.006} {\bibfield  {journal} {\bibinfo  {journal} {Eur. J. Comb.}\ }\textbf {\bibinfo {volume} {29}},\ \bibinfo {pages} {1881–1893} (\bibinfo {year} {2008})}\BibitemShut {NoStop}%
\bibitem [{Note6()}]{Note6}%
  \BibitemOpen
  \bibinfo {note} {We observe that for small values of $p$ short logical operators attached to the boundary can appear, thus we take large enough values to suppress these.}\BibitemShut {Stop}%
\bibitem [{\citenamefont {Bravyi}\ \emph {et~al.}(2010)\citenamefont {Bravyi}, \citenamefont {Poulin},\ and\ \citenamefont {Terhal}}]{Bravyi_PRL2010}%
  \BibitemOpen
  \bibfield  {author} {\bibinfo {author} {\bibfnamefont {S.}~\bibnamefont {Bravyi}}, \bibinfo {author} {\bibfnamefont {D.}~\bibnamefont {Poulin}},\ and\ \bibinfo {author} {\bibfnamefont {B.}~\bibnamefont {Terhal}},\ }\bibfield  {title} {\bibinfo {title} {Tradeoffs for reliable quantum information storage in 2d systems},\ }\href {https://doi.org/10.1103/PhysRevLett.104.050503} {\bibfield  {journal} {\bibinfo  {journal} {Phys. Rev. Lett.}\ }\textbf {\bibinfo {volume} {104}},\ \bibinfo {pages} {050503} (\bibinfo {year} {2010})}\BibitemShut {NoStop}%
\bibitem [{Note7()}]{Note7}%
  \BibitemOpen
  \bibinfo {note} {While we have studied a particular rectangular patch under the substitution rules, more general regions are possible, and due to the lack of spatial symmetry non-identical codes will not have large coincident subgraphs.}\BibitemShut {Stop}%
\bibitem [{\citenamefont {Williamson}\ and\ \citenamefont {Baspin}(2023)}]{Williamson_Layer2023}%
  \BibitemOpen
  \bibfield  {author} {\bibinfo {author} {\bibfnamefont {D.~J.}\ \bibnamefont {Williamson}}\ and\ \bibinfo {author} {\bibfnamefont {N.}~\bibnamefont {Baspin}},\ }\href@noop {} {\bibinfo {title} {Layer codes}} (\bibinfo {year} {2023}),\ \Eprint {https://arxiv.org/abs/2309.16503} {arXiv:2309.16503 [quant-ph]} \BibitemShut {NoStop}%
\bibitem [{\citenamefont {Vidal}(2007)}]{Vidal_ER2007}%
  \BibitemOpen
  \bibfield  {author} {\bibinfo {author} {\bibfnamefont {G.}~\bibnamefont {Vidal}},\ }\bibfield  {title} {\bibinfo {title} {Entanglement renormalization},\ }\href {https://doi.org/10.1103/PhysRevLett.99.220405} {\bibfield  {journal} {\bibinfo  {journal} {Phys. Rev. Lett.}\ }\textbf {\bibinfo {volume} {99}},\ \bibinfo {pages} {220405} (\bibinfo {year} {2007})}\BibitemShut {NoStop}%
\bibitem [{\citenamefont {Haah}(2014)}]{Haah_PRB2014}%
  \BibitemOpen
  \bibfield  {author} {\bibinfo {author} {\bibfnamefont {J.}~\bibnamefont {Haah}},\ }\bibfield  {title} {\bibinfo {title} {Bifurcation in entanglement renormalization group flow of a gapped spin model},\ }\href {https://doi.org/10.1103/PhysRevB.89.075119} {\bibfield  {journal} {\bibinfo  {journal} {Phys. Rev. B}\ }\textbf {\bibinfo {volume} {89}},\ \bibinfo {pages} {075119} (\bibinfo {year} {2014})}\BibitemShut {NoStop}%
\bibitem [{\citenamefont {Dua}\ \emph {et~al.}(2020)\citenamefont {Dua}, \citenamefont {Sarkar}, \citenamefont {Williamson},\ and\ \citenamefont {Cheng}}]{Dua_PRR2020}%
  \BibitemOpen
  \bibfield  {author} {\bibinfo {author} {\bibfnamefont {A.}~\bibnamefont {Dua}}, \bibinfo {author} {\bibfnamefont {P.}~\bibnamefont {Sarkar}}, \bibinfo {author} {\bibfnamefont {D.~J.}\ \bibnamefont {Williamson}},\ and\ \bibinfo {author} {\bibfnamefont {M.}~\bibnamefont {Cheng}},\ }\bibfield  {title} {\bibinfo {title} {Bifurcating entanglement-renormalization group flows of fracton stabilizer models},\ }\href {https://doi.org/10.1103/PhysRevResearch.2.033021} {\bibfield  {journal} {\bibinfo  {journal} {Phys. Rev. Res.}\ }\textbf {\bibinfo {volume} {2}},\ \bibinfo {pages} {033021} (\bibinfo {year} {2020})}\BibitemShut {NoStop}%
\bibitem [{\citenamefont {San~Miguel}\ \emph {et~al.}(2021)\citenamefont {San~Miguel}, \citenamefont {Dua},\ and\ \citenamefont {Williamson}}]{Dua_PRB2021}%
  \BibitemOpen
  \bibfield  {author} {\bibinfo {author} {\bibfnamefont {J.~F.}\ \bibnamefont {San~Miguel}}, \bibinfo {author} {\bibfnamefont {A.}~\bibnamefont {Dua}},\ and\ \bibinfo {author} {\bibfnamefont {D.~J.}\ \bibnamefont {Williamson}},\ }\bibfield  {title} {\bibinfo {title} {Bifurcating subsystem symmetric entanglement renormalization in two dimensions},\ }\href {https://doi.org/10.1103/PhysRevB.103.035148} {\bibfield  {journal} {\bibinfo  {journal} {Phys. Rev. B}\ }\textbf {\bibinfo {volume} {103}},\ \bibinfo {pages} {035148} (\bibinfo {year} {2021})}\BibitemShut {NoStop}%
\bibitem [{\citenamefont {Aitchison}\ \emph {et~al.}(2023)\citenamefont {Aitchison}, \citenamefont {Bulmash}, \citenamefont {Dua}, \citenamefont {Doherty},\ and\ \citenamefont {Williamson}}]{Aitchison_Strings2023}%
  \BibitemOpen
  \bibfield  {author} {\bibinfo {author} {\bibfnamefont {C.~T.}\ \bibnamefont {Aitchison}}, \bibinfo {author} {\bibfnamefont {D.}~\bibnamefont {Bulmash}}, \bibinfo {author} {\bibfnamefont {A.}~\bibnamefont {Dua}}, \bibinfo {author} {\bibfnamefont {A.~C.}\ \bibnamefont {Doherty}},\ and\ \bibinfo {author} {\bibfnamefont {D.~J.}\ \bibnamefont {Williamson}},\ }\href@noop {} {\bibinfo {title} {No strings attached: Boundaries and defects in the cubic code}} (\bibinfo {year} {2023}),\ \Eprint {https://arxiv.org/abs/2308.00138} {arXiv:2308.00138 [quant-ph]} \BibitemShut {NoStop}%
\bibitem [{\citenamefont {Li}\ and\ \citenamefont {Ye}(2020)}]{LiYe20}%
  \BibitemOpen
  \bibfield  {author} {\bibinfo {author} {\bibfnamefont {M.-Y.}\ \bibnamefont {Li}}\ and\ \bibinfo {author} {\bibfnamefont {P.}~\bibnamefont {Ye}},\ }\bibfield  {title} {\bibinfo {title} {Fracton physics of spatially extended excitations},\ }\href {https://doi.org/10.1103/PhysRevB.101.245134} {\bibfield  {journal} {\bibinfo  {journal} {Phys. Rev. B}\ }\textbf {\bibinfo {volume} {101}},\ \bibinfo {pages} {245134} (\bibinfo {year} {2020})}\BibitemShut {NoStop}%
\bibitem [{\citenamefont {Kogut}(1979)}]{Kogut_RMP1979}%
  \BibitemOpen
  \bibfield  {author} {\bibinfo {author} {\bibfnamefont {J.~B.}\ \bibnamefont {Kogut}},\ }\bibfield  {title} {\bibinfo {title} {An introduction to lattice gauge theory and spin systems},\ }\href {https://doi.org/10.1103/RevModPhys.51.659} {\bibfield  {journal} {\bibinfo  {journal} {Rev. Mod. Phys.}\ }\textbf {\bibinfo {volume} {51}},\ \bibinfo {pages} {659} (\bibinfo {year} {1979})}\BibitemShut {NoStop}%
\bibitem [{\citenamefont {Fradkin}\ and\ \citenamefont {Shenker}(1979)}]{Fradkin_PRD1979}%
  \BibitemOpen
  \bibfield  {author} {\bibinfo {author} {\bibfnamefont {E.}~\bibnamefont {Fradkin}}\ and\ \bibinfo {author} {\bibfnamefont {S.~H.}\ \bibnamefont {Shenker}},\ }\bibfield  {title} {\bibinfo {title} {Phase diagrams of lattice gauge theories with {H}iggs fields},\ }\href {https://doi.org/10.1103/PhysRevD.19.3682} {\bibfield  {journal} {\bibinfo  {journal} {Phys. Rev. D}\ }\textbf {\bibinfo {volume} {19}},\ \bibinfo {pages} {3682} (\bibinfo {year} {1979})}\BibitemShut {NoStop}%
\bibitem [{\citenamefont {Bollob{\'a}s}\ and\ \citenamefont {Klee}(1984)}]{Bollobas_Diameters1984}%
  \BibitemOpen
  \bibfield  {author} {\bibinfo {author} {\bibfnamefont {B.}~\bibnamefont {Bollob{\'a}s}}\ and\ \bibinfo {author} {\bibfnamefont {V.}~\bibnamefont {Klee}},\ }\bibfield  {title} {\bibinfo {title} {Diameters of random bipartite graphs},\ }\href@noop {} {\bibfield  {journal} {\bibinfo  {journal} {Combinatorica}\ }\textbf {\bibinfo {volume} {4}},\ \bibinfo {pages} {7} (\bibinfo {year} {1984})}\BibitemShut {NoStop}%
\bibitem [{\citenamefont {Chung}\ and\ \citenamefont {Lu}(2001)}]{Chung_Diameter2001}%
  \BibitemOpen
  \bibfield  {author} {\bibinfo {author} {\bibfnamefont {F.}~\bibnamefont {Chung}}\ and\ \bibinfo {author} {\bibfnamefont {L.}~\bibnamefont {Lu}},\ }\bibfield  {title} {\bibinfo {title} {The diameter of sparse random graphs},\ }\href {https://doi.org/https://doi.org/10.1006/aama.2001.0720} {\bibfield  {journal} {\bibinfo  {journal} {Advances in Applied Mathematics}\ }\textbf {\bibinfo {volume} {26}},\ \bibinfo {pages} {257} (\bibinfo {year} {2001})}\BibitemShut {NoStop}%
\bibitem [{\citenamefont {Newman}(2003)}]{Newman_SIAM2003}%
  \BibitemOpen
  \bibfield  {author} {\bibinfo {author} {\bibfnamefont {M.~E.~J.}\ \bibnamefont {Newman}},\ }\bibfield  {title} {\bibinfo {title} {The structure and function of complex networks},\ }\href {https://doi.org/10.1137/S003614450342480} {\bibfield  {journal} {\bibinfo  {journal} {SIAM Review}\ }\textbf {\bibinfo {volume} {45}},\ \bibinfo {pages} {167} (\bibinfo {year} {2003})}\BibitemShut {NoStop}%
\bibitem [{\citenamefont {Grospellier}\ \emph {et~al.}(2021)\citenamefont {Grospellier}, \citenamefont {Grou{\`{e}}s}, \citenamefont {Krishna},\ and\ \citenamefont {Leverrier}}]{Grospellier_quantum2021}%
  \BibitemOpen
  \bibfield  {author} {\bibinfo {author} {\bibfnamefont {A.}~\bibnamefont {Grospellier}}, \bibinfo {author} {\bibfnamefont {L.}~\bibnamefont {Grou{\`{e}}s}}, \bibinfo {author} {\bibfnamefont {A.}~\bibnamefont {Krishna}},\ and\ \bibinfo {author} {\bibfnamefont {A.}~\bibnamefont {Leverrier}},\ }\bibfield  {title} {\bibinfo {title} {Combining hard and soft decoders for hypergraph product codes},\ }\href {https://doi.org/10.22331/q-2021-04-15-432} {\bibfield  {journal} {\bibinfo  {journal} {{Quantum}}\ }\textbf {\bibinfo {volume} {5}},\ \bibinfo {pages} {432} (\bibinfo {year} {2021})}\BibitemShut {NoStop}%
\bibitem [{\citenamefont {Tremblay}\ \emph {et~al.}(2022)\citenamefont {Tremblay}, \citenamefont {Delfosse},\ and\ \citenamefont {Beverland}}]{Tremblay_PRL2022}%
  \BibitemOpen
  \bibfield  {author} {\bibinfo {author} {\bibfnamefont {M.~A.}\ \bibnamefont {Tremblay}}, \bibinfo {author} {\bibfnamefont {N.}~\bibnamefont {Delfosse}},\ and\ \bibinfo {author} {\bibfnamefont {M.~E.}\ \bibnamefont {Beverland}},\ }\bibfield  {title} {\bibinfo {title} {Constant-overhead quantum error correction with thin planar connectivity},\ }\href {https://doi.org/10.1103/PhysRevLett.129.050504} {\bibfield  {journal} {\bibinfo  {journal} {Phys. Rev. Lett.}\ }\textbf {\bibinfo {volume} {129}},\ \bibinfo {pages} {050504} (\bibinfo {year} {2022})}\BibitemShut {NoStop}%
\bibitem [{\citenamefont {Cooperman}\ \emph {et~al.}(1999)\citenamefont {Cooperman}, \citenamefont {Feisel}, \citenamefont {von~zur Gathen},\ and\ \citenamefont {Havas}}]{Cooperman_GCD1999}%
  \BibitemOpen
  \bibfield  {author} {\bibinfo {author} {\bibfnamefont {G.}~\bibnamefont {Cooperman}}, \bibinfo {author} {\bibfnamefont {S.}~\bibnamefont {Feisel}}, \bibinfo {author} {\bibfnamefont {J.}~\bibnamefont {von~zur Gathen}},\ and\ \bibinfo {author} {\bibfnamefont {G.}~\bibnamefont {Havas}},\ }\bibfield  {title} {\bibinfo {title} {{GCD} of many integers},\ }in\ \href@noop {} {\emph {\bibinfo {booktitle} {International Computing and Combinatorics Conference}}}\ (\bibinfo {organization} {Springer},\ \bibinfo {year} {1999})\ pp.\ \bibinfo {pages} {310--317}\BibitemShut {NoStop}%
\bibitem [{Note8()}]{Note8}%
  \BibitemOpen
  \bibinfo {note} {Another way to arrange the tile aperiodically is via its matching rules~\cite {Radin_1994Pinwheel}, which are not employed here but allow for more general boundaries.}\BibitemShut {Stop}%
\bibitem [{\citenamefont {Reed}\ and\ \citenamefont {Solomon}(1960)}]{Reed_SIAM1960}%
  \BibitemOpen
  \bibfield  {author} {\bibinfo {author} {\bibfnamefont {I.~S.}\ \bibnamefont {Reed}}\ and\ \bibinfo {author} {\bibfnamefont {G.}~\bibnamefont {Solomon}},\ }\bibfield  {title} {\bibinfo {title} {Polynomial codes over certain finite fields},\ }\href {https://doi.org/10.1137/0108018} {\bibfield  {journal} {\bibinfo  {journal} {Journal of the society for industrial and applied mathematics}\ }\textbf {\bibinfo {volume} {8}},\ \bibinfo {pages} {300} (\bibinfo {year} {1960})}\BibitemShut {NoStop}%
\bibitem [{\citenamefont {Cancellieri}(2015)}]{Cancellieri_Polynomial2015}%
  \BibitemOpen
  \bibfield  {author} {\bibinfo {author} {\bibfnamefont {G.}~\bibnamefont {Cancellieri}},\ }\href {https://doi.org/10.1007/978-3-319-01727-3} {\emph {\bibinfo {title} {Polynomial theory of error correcting codes}}}\ (\bibinfo  {publisher} {Springer},\ \bibinfo {year} {2015})\BibitemShut {NoStop}%
\bibitem [{\citenamefont {Kubica}\ \emph {et~al.}(2015)\citenamefont {Kubica}, \citenamefont {Yoshida},\ and\ \citenamefont {Pastawski}}]{Kubica_2015}%
  \BibitemOpen
  \bibfield  {author} {\bibinfo {author} {\bibfnamefont {A.}~\bibnamefont {Kubica}}, \bibinfo {author} {\bibfnamefont {B.}~\bibnamefont {Yoshida}},\ and\ \bibinfo {author} {\bibfnamefont {F.}~\bibnamefont {Pastawski}},\ }\bibfield  {title} {\bibinfo {title} {Unfolding the color code},\ }\href {https://doi.org/10.1088/1367-2630/17/8/083026} {\bibfield  {journal} {\bibinfo  {journal} {New Journal of Physics}\ }\textbf {\bibinfo {volume} {17}},\ \bibinfo {pages} {083026} (\bibinfo {year} {2015})}\BibitemShut {NoStop}%
\bibitem [{\citenamefont {Shirley}\ \emph {et~al.}(2019{\natexlab{b}})\citenamefont {Shirley}, \citenamefont {Slagle},\ and\ \citenamefont {Chen}}]{Shirley19}%
  \BibitemOpen
  \bibfield  {author} {\bibinfo {author} {\bibfnamefont {W.}~\bibnamefont {Shirley}}, \bibinfo {author} {\bibfnamefont {K.}~\bibnamefont {Slagle}},\ and\ \bibinfo {author} {\bibfnamefont {X.}~\bibnamefont {Chen}},\ }\bibfield  {title} {\bibinfo {title} {Foliated fracton order in the checkerboard model},\ }\href {https://doi.org/10.1103/PhysRevB.99.115123} {\bibfield  {journal} {\bibinfo  {journal} {Phys. Rev. B}\ }\textbf {\bibinfo {volume} {99}},\ \bibinfo {pages} {115123} (\bibinfo {year} {2019}{\natexlab{b}})}\BibitemShut {NoStop}%
\bibitem [{\citenamefont {Bravyi}\ and\ \citenamefont {Hastings}(2014)}]{Bravyi_ACM2014}%
  \BibitemOpen
  \bibfield  {author} {\bibinfo {author} {\bibfnamefont {S.}~\bibnamefont {Bravyi}}\ and\ \bibinfo {author} {\bibfnamefont {M.~B.}\ \bibnamefont {Hastings}},\ }\bibfield  {title} {\bibinfo {title} {Homological product codes},\ }in\ \href {https://doi.org/10.1145/2591796.2591870} {\emph {\bibinfo {booktitle} {Proceedings of the Forty-Sixth Annual ACM Symposium on Theory of Computing}}},\ \bibinfo {series and number} {STOC '14}\ (\bibinfo  {publisher} {Association for Computing Machinery},\ \bibinfo {address} {New York, NY, USA},\ \bibinfo {year} {2014})\ p.\ \bibinfo {pages} {273–282}\BibitemShut {NoStop}%
\bibitem [{\citenamefont {Radicevic}(2020)}]{Radicevic_Arxiv2020}%
  \BibitemOpen
  \bibfield  {author} {\bibinfo {author} {\bibfnamefont {D.}~\bibnamefont {Radicevic}},\ }\href@noop {} {\bibinfo {title} {Systematic constructions of fracton theories}} (\bibinfo {year} {2020}),\ \Eprint {https://arxiv.org/abs/1910.06336} {arXiv:1910.06336 [cond-mat.str-el]} \BibitemShut {NoStop}%
\end{thebibliography}%

\clearpage
\onecolumngrid

\setcounter{equation}{0}
\setcounter{figure}{0}
\setcounter{table}{0}
\setcounter{page}{1}
\makeatletter

\begin{center}
{\textbf{\Large Supplemental material: Fracton models from product codes}}\\
(Dated: \today)
\end{center}

\section{Conditions for fractons in product codes}
\subsection{Rank deficiency}
\subsubsection{Rank deficiency in generic Tanner graphs}

Consider a classical code specified by a Tanner graph with $n$ bit nodes and $m$ check nodes.
Denoting the bit and check vector spaces over $\F_2$ as $\V_b$ and $\V_c$, respectively, the parity check matrix $H: \V_b \to \V_c$ is a map taking error configurations $e \in \V_b$ to syndromes $s \in \V_c$.
One can define a classical Hamiltonian $\mathcal H_\mathrm{cl} = -\sum_{i=1}^m H_i$, where $i$ is the row index of the parity check matrix; that is, the $H_i$ are the checks.
Ground states of $\mathcal H_\mathrm{cl}$ form the preimage of the zero-syndrome configuration $s_\mathrm{vac}=0 \in \V_c$.
Assuming no syndrome measurement error, a physical error channel produces only syndromes of the form $s_\mathrm{vac} + H\cdot e$ for some error configuration $e \in \V_b$.

If $m>n$ the parity check matrix is not full-rank and $\img(H)\subseteq \V_c$ is strict, meaning that syndrome configurations exist which are not connected to physical ground states.
The quotient $\V_c/\img(H)$ contains such ``unphysical'' syndromes in the classical code, and the codimension $k^\top = m - \rank(H)$ counts the cosets.
But this is nothing but the number of logical bits in the transpose (or dual) code $H^\top$.
This relationship between unphysical syndromes and logical bits is a feature of the duality of classical codes.

\subsubsection{From rank deficiency to superselection sectors}

In a stabilizer Hamiltonian, superselection sectors are equivalence classes of simple or composite excitations.
Intuitively, they describe what remains within a region after a process in which particles maximally annihilate each other.
In CSS quantum codes, $X$ and $Z$ stabilizers are treated using independent parity check matrices $H_X$ and $H_Z$.
Consequently, superselection sectors can be labeled as either $X$ or $Z$ type, and the previous discussion of classical parity check matrices can be applied.
For a HGP code, the number of superselection sectors is given by \cite{Tillich_TIT2014}
\begin{equation}
    k_X^\top = k_1^\top k_2~,\qquad k_Z^\top = k_1 k_2^\top.
\end{equation}
In particular, the superselection sectors form a group $\Z_2^{k_X^\top}\times \Z_2^{k_Z^\top}$.
In the case of the toric code, $k_X^\top = k_Z^\top = 1$, and the group is the anyon theory $\{1,e,m,\psi\}$.
The more codewords the classical codes (or their duals) have, the larger the number of superselection sectors in the quantum code.
One can also see the connection of superselection sectors to the number of logical qubits in HGP quantum codes ($n_q=n_1 n_2+m_1 m_2$):
\begin{equation}
        k_q = k_X + k_Z - n_q 
            = (m_X - n_q + k_X ^\top) + (m_Z - n_q + k_Z ^\top) - n_q
            = k_1 k_2 + k_1^\top k_2^\top~.
\end{equation}

\subsection{Confinement}

We call a classical code \emph{confining} if all error vectors $e \in \V_b$ with Hamming weight $|e| \leq g(n)$ satisfy $f(|s|) \geq |e|$, where $s = H\cdot e$, for some increasing functions $f$ and $g$~\cite{Quintavalle_PRX2021}.
Evidently $g$ cannot grow faster than the code distance $d$.
This criterion is weaker than another standard usage of ``confinement'' \cite{Bombin_PRX2015}: for example, the signature of the deconfined phase of a gauge theory is perimeter law scaling of Wilson loops, whereas a confined phase is characterized by an area law \cite{Kogut_RMP1979}.
Both of these cases would be considered confining under the present definition, with the area law satisfying the stronger condition of \emph{linear confinement}.
In this language, stabilizer codes are not generically confining on a lattice because they realize the pure gauge limit of such field theories, and away from this limit confinement is a result of coupling to gapped matter fields~\cite{Fradkin_PRD1979}.
There is a tradeoff between confinement and encoding properties: for example, the trivial stabilizer code $\mathcal H = -\sum_j Z_j$ is linearly confining, but lacks logical qubits.

By contrast, a ``no-strings'' condition is inconvenient for characterizing nonlocal codes because it relies on additional structure, including a separation of scales and thus a metric.
Essentially, it stipulates that collections of excitations within a finite region of characteristic length $r$ cannot be moved a distance $\ell \gg r$ by an $O(\ell)$ error.
One concrete such condition for a general LDPC code is as follows.
We note that the Tanner graph becomes a metric space when endowed with a distance function (which may be graph distance, for example, or arise from an underlying manifold).
We say that the code contains no strings of length $\ell$ if for all error configurations $e_\mathrm{ns}(\ell) \in \V_b$ such that $H \cdot e_\mathrm{ns} = \sigma+\sigma'$, where $\sigma,\sigma' \in \V_c$ satisfy the conditions below, it is the case that $|e_\mathrm{ns}(\ell)| \geq h(\ell)$ for some super-linear polynomial $h$.
We place the following conditions on the syndromes $\sigma$ and $\sigma'$:
\begin{enumerate}
\item nontrivial support only within an $r$-ball $\B_r$ for some constant $r$;
\item no preimage $\tilde e$, $\sigma = H \cdot \tilde e$, where $|\tilde e|$ is bounded by a constant;
\end{enumerate}
additionally, the minimum distance between $\B_r$ (corresponding to $\sigma$) and $\B_{r'}$ (corresponding to $\sigma'$) is $\ell$.

Confinement implies no strings as defined above: evidently $|H\cdot e_\mathrm{ns}|$ is bounded independent of $\ell$, so $|e_\mathrm{ns}|$ is either bounded, or grows faster than $g(n)$.
Both of these contradict the definition of a string operator.

\subsection{Isolability}

Isolability ensures that excitations are fundamentally pointlike, rather than comprising extended objects like the domain walls of the Ising model in two or more dimensions.
As the technique is to ensure that the distance between excitations can be made large, a metric is required, for which we employ the graph distance.
Consider an excitation $\sigma \in \V_c$ which satisfies the same conditions described for the syndromes in the no-strings condition above.
Denote by $\Lambda$ the subgraph consisting of all $R$-neighbors of $\mathcal B_r$, where $|\Lambda| \leq \kappa^R$.
The distance $R=w(n)$ should be taken to be some increasing function of $n$.
Isolability requires that $\img(H)$ contains a vector $\sigma + \overline\sigma$, where $\overline\sigma$ has support only on the complement of $\Lambda$.

This condition can be understood in several limits: for instance, almost all random bipartite graphs without sparsity conditions (i.e., not LDPC) have diameter bounded by 4, so are not candidates for isolability~\cite{Bollobas_Diameters1984}.
Conversely, a tree Tanner graph exhibits $R\sim\log n$ isolability.
As a coarse measure of the naturalness of isolability for typical LDPC codes, the dimension of the error space $n$ can be compared to the number of free parameters in such an isolation vector, namely $m-\kappa^R$.
As $m$ scales linearly with $n$, one naively expects $R\sim\log n$ isolability, as in the case of the tree.
In fact, because the diameter of a random sparse graph is logarithmic in $n$ with high probability, this scaling is optimal~\cite{Chung_Diameter2001}.

\begin{figure}[ht]
\centering
\includegraphics[width=0.25\columnwidth]{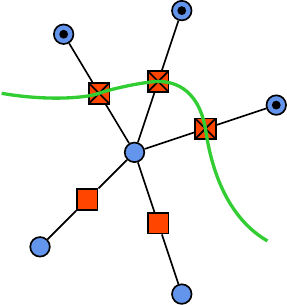}\hspace{3cm}
\includegraphics[width=0.25\columnwidth]{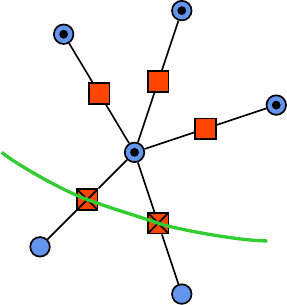}
\caption{Lack of isolability is illustrated as a consequence of a local symmetry.
Errors (denoted by black dots) preserve the domain-wall nature of the excitation (denoted by crosses) by implementing simple deformations pulling the green surface through a bit.
This feature is not generic on Tanner graphs with $\kappa_c > 2$.}
\label{fig:iso_cond}
\end{figure}

In the main text we identify a necessary condition for a Tanner graph to be isolable, namely the lack of ``Ising subgraphs,'' regions consisting entirely of checks of valence two.
Within these features excitations manifest as deformable domain walls surrounding regions of errors.
This property is quite special, and arises from a local symmetry which allows a domain wall to be pulled through a bit by applying an error, as illustrated in Fig.~\ref{fig:iso_cond}.
There is a specific type of local \emph{meta-check} associated to each independent cycle within the Ising subgraph.
(Consequently, cyclic or tree-like Ising subgraphs which have only one or no associated meta-checks actually remain isolable.
While these have vanishing probability even compared to typical Ising subgraphs in a nonlocal code, in the local case, the Ising model on a cycle---that is, a cyclic repetition code---does not spoil isolability.)
Outside of Ising subgraphs, excitations are no longer necessarily connected in describing such a boundary; that is, an excitation pattern can itself develop a boundary and thus become disconnected, leading to isolability in the typical case.

\section{Type-I fractons on graphs}

In each of the following constructions, we probabilistically generate either connectivity graphs or Tanner graphs which are used to underlie nonlocal codes.
In each case we wish to sample fairly from sparse graphs, without biasing toward any particular additional structure.
One requirement we do impose, however, is that the resulting graphs must be connected.
This is in order to avoid trivial instances of quantum codes such as stacks of toric codes.
We note that a hypergraph product is connected iff its seed codes are connected.

\subsection{Laplacian codes}\label{sec:laplacian_code}

The Laplacian matrix of an arbitrary undirected graph $\Gamma$ with vertices $\{v_i,\ldots, v_n\}$ is given by $L = D-A$, where $D$ is a diagonal matrix whose element $D_{ii}$ is the degree of vertex $v_i$; and $A$ is the adjacency matrix, meaning that $A_{ij}=1$ if an edge connects $v_i$ and $v_j$ and $A_{ij}=0$ otherwise~\cite{Chung_1997}.
The classical Laplacian code on a graph $\Gamma$ has parity check matrix given by the reduction mod 2 of the Laplacian: $H = L \mod 2$.
Each vertex $v_i$ then represents both a bit and a check, and $\Gamma$ provides the connectivity graph.
The code obtained this way is symmetric (equivalently, self-dual): $H = H^\top$, with diagonal elements being equal to the parity of the off-diagonal elements in each row or column.
Consequently each row and column sums to 0.

\subsubsection{Configuration model}
To construct an ensemble of typical Laplacian codes, we employ an algorithm called the \emph{configuration model} which generates random graphs with a given degree sequence ${\mathcal{D}_1,\ldots, \mathcal{D}_n}$.
This method is heavily employed in the LDPC literature to generate randomized codes while allowing for convenient control of degrees of vertices \cite{Newman_SIAM2003, Grospellier_quantum2021, Tremblay_PRL2022}.
We set a constant upper bound $\mathcal{D}_{\text{high}}=5$, which guarantees the code we construct is LDPC; we also set a lower bound $\mathcal{D}_{\text{low}}=3$ to completely avoid Ising subgraphs.
The resulting graph from the configuration model may contain contain self-loops or parallel edges between the same pair of vertices; we discard these redundant connections.
In the main text an ensemble of $10^3$ random graphs is generated following the above procedure, which constitutes a measure useful for studying rank deficiency and confinement of Laplacian codes.

\begin{figure}[ht]
\centering
\includegraphics[width=0.7\columnwidth]{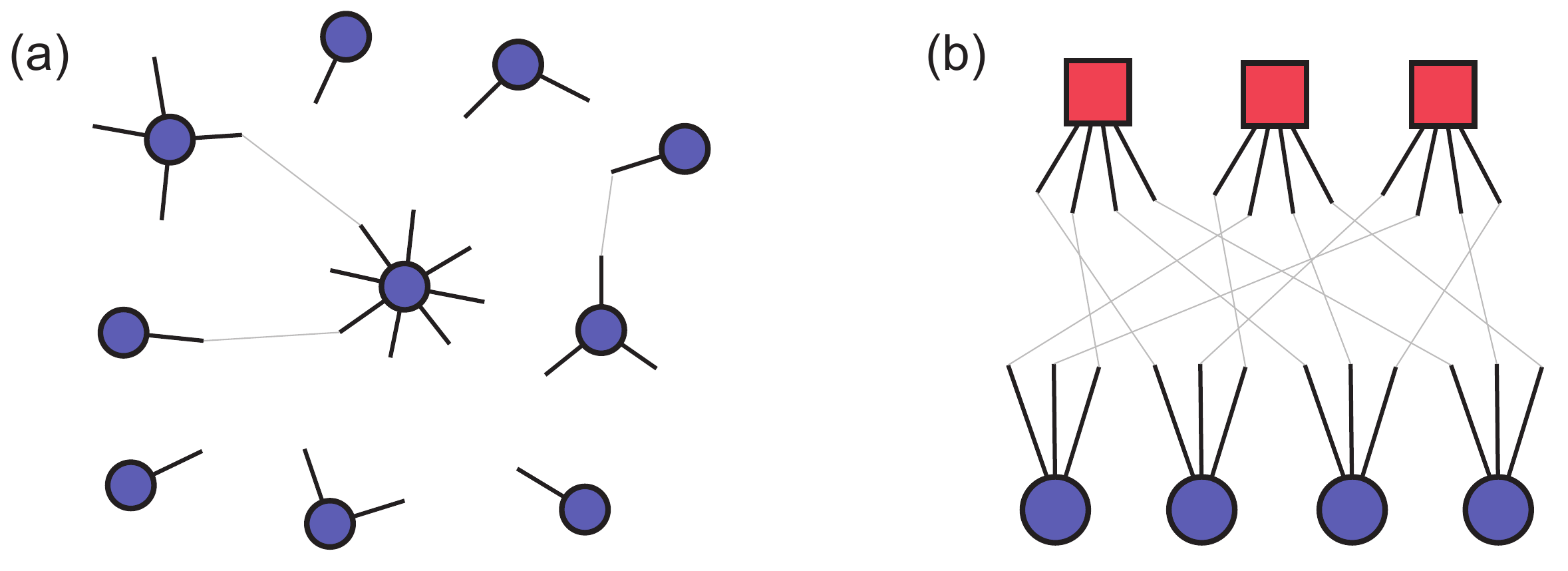}
\caption{\label{fig:configuration_model}
Configuration models are illustrated.
(a) In the case of typical Laplacian codes, to construct random graphs with bounded degree, one generates a sequence of nodes carrying a given number of ``half-edges.''
Half-edges from different nodes are randomly paired to form a connected graph, with parallel edges and self-loops removed in the end.
(b) In the case of typical LDPC codes, the bipartite Tanner graph is constructed; here red squares represent check nodes and blue circles represent variable nodes.
Only half-edges carried by different species of nodes are paired.
Assuming regularity, to ensure complete pairing of half-edges, the number of check nodes $m$, the degree of check nodes $\mathcal{D}_{\text{check}}$, the number of variable nodes $n$, and the degree of variable nodes $\mathcal{D}_{\text{variable}}$, should satisfy $m \mathcal{D}_{\text{check}}=n \mathcal{D}_{\text{variable}}$.
}
\end{figure}

\subsubsection{Rank of Laplacian codes}
\label{sec:laplacian_rank}

The rank of a graph Laplacian is given by the number of vertices, with the number of connected components subtracted.
If $\Gamma$ is connected, the emergence of many logicals in the classical code is due to the reduction mod 2.
Specifically, $k$ is the sum of the reduction mod 2 of the invariant factors of $\Gamma$ \cite{Gorantla_PRB2023,Ebisu_Scipost2023}.
A global spin flip always contributes a single logical bit, regardless of the graph.
The other encoded bits rely on the graph; for example, the Laplacian code on a cycle graph $\Gamma=C_n$ supports one additional logical bit with a N{\'e}el-type codeword for even $n$, but none for odd $n$.
The complete graph $\Gamma=K_n$ has $n-2$ additional logical bits if $n$ is even and again none if $n$ is odd.
Because the product of invariant factors is the number of spanning trees, the Laplacian code on a tree has only the single logical bit.

Using the detailed results of Refs.~\cite{Gorantla_PRB2023,Ebisu_Scipost2023} we can understand how translation invariance leads to an enhanced codespace in the Laplacian code.
Namely, rank deficiency requires that many of the invariant factors (over $\Z$) of $\Gamma$ be even.
These invariant factors are computed from the greatest common divisor of the determinants of certain minors of the Laplacian matrix.
On a graph without a high degree of structure, one may naively treat the determinants as random integers; then the probability that the gcd of $\mathcal N$ of these is equal to 1 is $\frac{1}{\zeta(\mathcal N)}$ (where $\zeta$ is the Riemann zeta function), which very rapidly approaches unity for large $\mathcal N$ \cite{Cooperman_GCD1999}.
In other words, a degree of approximate translational symmetry leading to correlations in the determinants of the minors of the Laplacian is necessary in order for the code to become rank deficient.
This evidently describes both the square lattice and the complete graph, two concrete instances for which the code exhibits a large number of encoded bits.

\subsubsection{Lack of confinement in Laplacian codes on regular lattices}
\label{sec:laplacian_confinement}

\begin{figure}
\centering
\includegraphics[width=0.75\columnwidth]{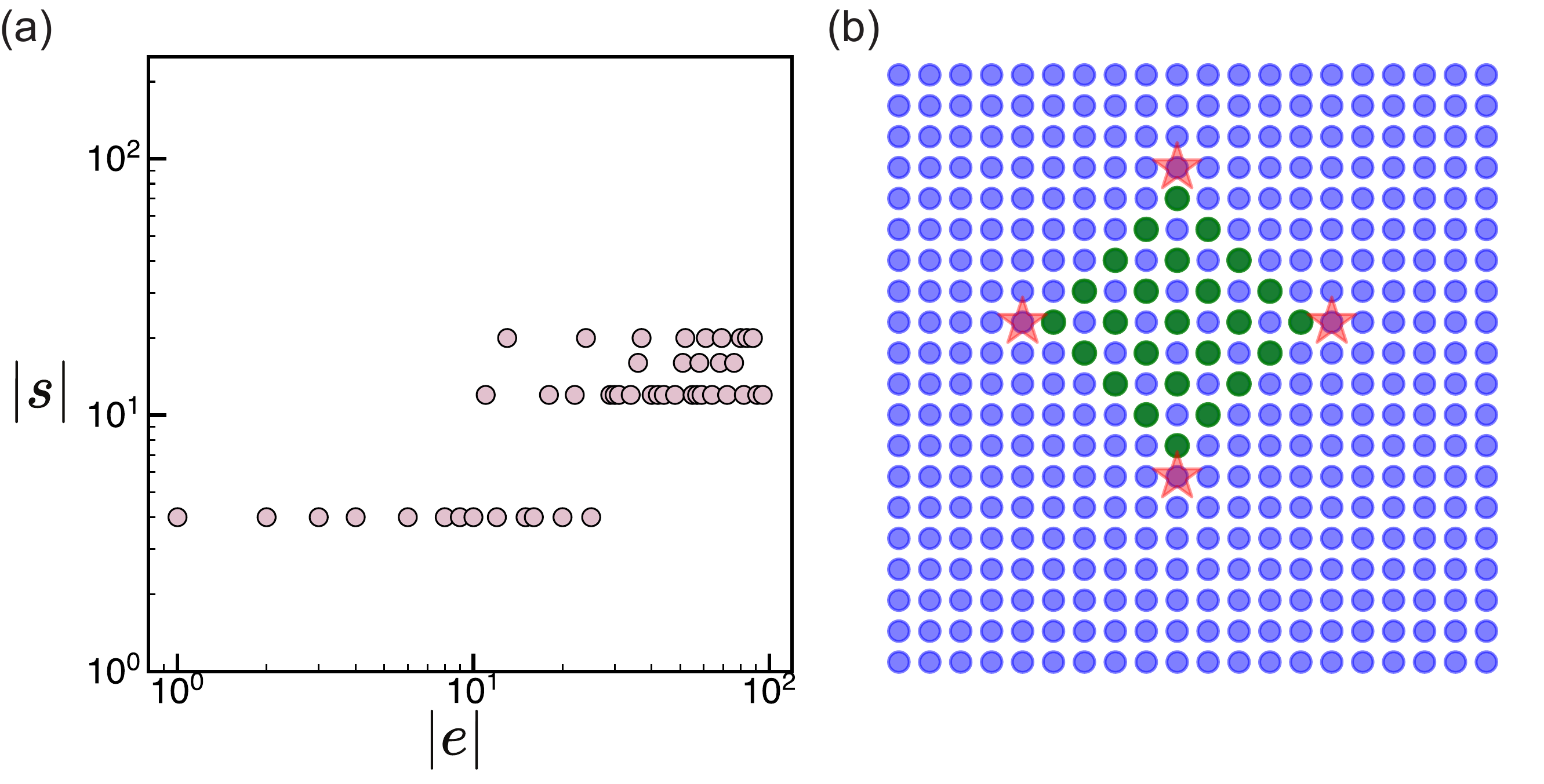}
\caption{\label{fig:laplacian_confinement}
Lack of confinement of the Laplacian code on the two-dimensional torus is shown.
(a) The confinement test via biased sampling over error configurations is performed for a $20\times 20$ square lattice.
Though we do not perform an exhaustive search at each weight, we use information from the logical operators in order to suggest low-energy errors, and in this case this strategy is sufficient to indicate the lack of scaling of the tails of the distribution, as arbitrarily large errors can have a syndrome weight of four.
(b) An example large error configuration having a syndrome of only weight four which was captured by the biased sampling procedure is exhibited.
}
\end{figure}

In the Laplacian code on the two-dimensional square lattice, a checkerboard pattern of errors in a rectangular region can generate excitations only at the corners, so has bounded syndrome weight of four.
The Laplacian code is thus not confining; indeed, it is not expected to be, as confinement is not natural on a regular lattice.
We provide this case as an example to justify the sampling procedure used in the main text to diagnose confinement in nonlocal and local codes.
Fig.~\ref{fig:laplacian_confinement} shows the result of the sampling for this case, in which the lower bound on syndrome weight is not monotonically increasing but instead fixed.

In order to consider errors with low syndrome weight, we bias the sampling toward energetically favorable configurations by choosing local truncations of logical operators.
That is, we choose errors by randomly selecting a base point in the system and a codeword of minimal weight, then placing only errors which appear in the codeword and are within some distance of the base point.
While only a heuristic, this protocol does prove to be more effective at finding low-weight errors than unbiased sampling.
For example, if a system has string operators which are allowed to attach to the boundaries, this protocol will identify truncated strings as low-weight errors, whereas random sampling of error configurations is less likely to do so.

\subsection{Typical LDPC codes}

Generic LDPC codes differ from Laplacian codes in two significant ways.
First, for generic LDPC codes it is conventional to choose the number of bits $n$ to be larger than the number of checks $m$.
The parity check matrix $H$ thus represents an under-determined linear system with fewer constraints (checks) than variables (bits).
Second, generic LDPC codes admit only a more general Tanner graph representation, which is a bipartite graph composed of two different species of nodes.
In Fig.~\ref{fig:configuration_model}(b), one species of nodes represents checks, depicted as squares, and the other species represents variable nodes, depicted as circles.
Typical LDPC codes are again constructed using configuration model, but with the constraint of being bipartite: each check node connects its half-edges only to the half-edges of variable nodes, and vice versa.

The fact that typical LDPC codes are underdetermined linear systems conveniently introduces rank deficiency.
In particular, when constructing LDPC codes using configuration model, it is conventional to set the number of checks $m$ and number of variables $n$ by setting the degree of checks $\mathcal{D}_{\text{check}}$ and variables $\mathcal{D}_{\text{variable}}$ in the Tanner graph.
For the number of half-edges from check nodes and variable nodes to match, one must have $m \mathcal{D}_{\text{check}}=n \mathcal{D}_{\text{variable}}$.
This way, any constant difference between the degrees of check and variable nodes will translate into a lower bound on rank deficiency which is linear in $n$.

\section{Product codes for local fractons}

\subsection{Conditions for classical seed codes}

The construction of classical seed codes for fractons subject to geometric locality is more technical than the $\kappa$-local case.
Based on the preceding discussion, a natural first choice is the Laplacian code on the two-dimensional square lattice.
While this code is indeed rank deficient with periodic boundary conditions due to meta-checks---nontrivial consistency conditions on the local terms---it also hosts associated string operators in the bulk.
These strings are precisely aligned with certain lattice vectors and are not deformable in the transverse direction (similar to the two-dimensional Ising plaquette model~\cite{Vijay_PRB2016}); nevertheless excitations observe constrained propagation without energy penalty, so this code is suitable only for Type-I fractons.

There is thus a tension in this code between the objectives of rank deficiency and confinement: whereas meta-checks are responsible for rank deficiency, they lead directly to string operators.
The resolution explored here is to turn to a different source of codewords, more in keeping with the nonlocal constructions above.
Namely, we force logical bits by introducing an imbalance between physical bits and checks, a quantity which lower bounds $k$.
However, for local interactions, this condition becomes rather delicate: the reason is the classical BPT bound $k\sqrt d = O(n)$ for local two-dimensional codes~\cite{Bravyi_PRL2010}.
As a consequence, if the imbalance is linear in $n$---arising, say, from some average depletion of checks in the system---then the code distance is necessarily bounded by a constant.

\subsection{Aperiodic codes}

To accommodate the classical BPT bound we thus introduce open boundary conditions and delete checks associated with vertices on the edge.
This scheme of boundary depletion ensures that an imbalance develops between the number of bits and checks which scales to leading order as $\sqrt n$.
If local meta-checks do not appear in the bulk of the system which additionally contribute to $k$ a term linear in $n$, then the code distance may scale up to linearly with the volume.

It is thus desirable that the bulk of a code does not support meta-checks, which are intuitively a result of spatial symmetries (see also Sec.~\ref{sec:laplacian_rank}).
A natural set of candidates is graphs defined by aperiodic tilings.
While these include quasicrystals, this subclass retains discrete rotational symmetry about certain points in the lattice, which can in principle lead to meta-checks causing string operators or finite logicals.
We instead choose to work with the specific example of the ``pinwheel'' tiling, which has no discrete translation invariance~\cite{Radin_Space1992,Radin_1994Pinwheel,Radin_Aperiodic1997} as well as statistical circular symmetry~\cite{Frettloh_Substitution2008}.
The substitution rule defining this tiling, which takes one triangular face to five, naturally generates a family of planar graphs $\{G_N\}$ of exponentially increasing size with $N$, starting from an initial collection of base tiles~\footnote{Another way to arrange the tile aperiodically is via its matching rules~\cite{Radin_1994Pinwheel}, which are not employed here but allow for more general boundaries.}.

In order to ensure bulk self-duality of the pinwheel code, we place both bits and checks on the vertices of a graph $G_N$.
The parity check matrix is based on the graph Laplacian $L_N$~\cite{Chung_1997}, but with diagonal elements satisfying odd parity: $\tilde H_N = L_N - \I$, so that each row and column of $\tilde H_N$ sums to $1$.
This choice introduces frustration in the sense of preventing codewords arising from global (or large-scale coherent) spin flips, which appear in the code space of both the Ising model and the Laplacian code on a general graph.
The parity check matrix $H_N$ of the pinwheel code is obtained by boundary depletion of checks in $\tilde H_N$, ensuring a square-root lower bound on $k$ as a function of the volume.

Although the construction does not guarantee either linear scaling of the code distance or confinement, we find evidence for both, as described in the main text.
In principle boundaries must be chosen in such a way as to prevent short edge logicals; nevertheless we find that under the naive prescription of spacing depletion sites evenly and sufficiently far apart along the rectangular edge, we do not typically observe these.
In contrast, confinement is a property of the bulk and is very difficult to study numerically.
However, the procedure employed here of locally truncating low-weight logicals is sufficient to identify violations of confinement in many classical and quantum models including the Laplacian code (see Sec.~\ref{sec:laplacian_confinement}), the toric code, and Haah's code.
Intuitively, a code's lowest-weight logical operators may be expected to correspond to confinement-violating bulk errors which are allowed to ``condense'' on the system boundaries.
The observation that the pinwheel code distance scales linearly with the volume thus at least allows the possibility of a confining bulk.

\section{Relationships between fracton models and product codes}

\subsection{From polynomial formalism to parity check formalism}

The polynomial formalism or stabilizer map is a convenient way to study quantum codes by applying algebraic techniques \cite{Haah_Commuting2013,*Haah_Algebraic2016}.
It is descended from an analogous framework applied to classical codes~\cite{Reed_SIAM1960,Cancellieri_Polynomial2015}.
We briefly review the translation from polynomials to the parity check formalism, which is used throughout our work for both classical and quantum codes.
A polynomial $f(\bm x) \in \F_2[\bm x]$, $\bm x = (x_0,x_1,\ldots,x_D)$ is in one-to-one correspondence with an $L^D$-length protograph $\lambda_f$ having symmetry $(C_L)^D$, under the usual map \cite{Panteleev_ITI2022}.
The parity check matrix of a code based on such protographs has elements taken from the ring of circulants.
One obtains the $\F_2$-valued parity check matrix straightforwardly by expanding the protographs in the matrix representation.
The value of this representation lies in the bijection between polynomials and circulants, which establishes the natural connection to the lifted product (LP) construction \cite{Panteleev_ITI2022}.

\subsection{Product code constructions of known fracton models}

In this section we provide a case-by-case examination of correspondences between fractons and product codes.
Neither category is strictly more expressive than the other; for example, while the polynomial formalism captures non-CSS models, product codes need not exhibit lattice translation symmetry.
Many interesting models are contained in the intersection of the polynomial formalism and product codes: these include cubic codes~\cite{Haah_PRA2011} and fractal spin liquids~\cite{Yoshida_PRB2013} as well as the Type-I fracton models the checkerboard and $X$-cube models~\cite{Vijay_PRB2016}.

\subsubsection{Fractal spin liquids}
A prototypical class of stabilizer models hosting fractons is fractal spin liquids, defined on the 3D cubic lattice~\cite{Yoshida_PRB2013}.
This class includes cubic codes, the canonical example of which is Haah's code, a Type-II fracton model~\cite{Haah_PRA2011}.
We review the construction of Haah's code as a lifted product of classical codes~\cite{Panteleev_ITI2022,Panteleev_ACM2022}.
The generating maps for (the CSS instances of) such codes take the form
\begin{equation}
    \sigma=\left(\begin{array}{cc}
    f & 0\\
    g & 0\\
    \hline
    0 & \overline g\\
    0 & \overline f\\     
    \end{array}\right).
    \label{eq:cubic_fsl}
\end{equation}
This stabilizer map $\sigma$ describes a CSS code having a single $X$-type stabilizer and a single $Z$-type stabilizer on a cubic lattice with two qubits per site.
That is, the parity check matrices are $H_X = \begin{bmatrix}\lambda_f&\lambda_g\end{bmatrix}$ and $H_Z = \begin{bmatrix}\lambda_{\overline g}&\lambda_{\overline f}\end{bmatrix}$, where $\lambda_f$ and $\lambda_g$ are $(C_L)^3$ protographs defined by the polynomials $f$ and $g$, respectively, and $\lambda_{\overline f} = \lambda_f^\top$.
For example, Haah's code is given by $f = 1+x+y+z$ and $g = 1 + xy + yz + xz$~\cite{Haah_PRA2011}.

The LP of two protograph-valued classical parity check matrices $H_1$ of dimension $m_1 \times n_1$ and $H_2$ of dimension $m_2\times n_2$ is given by~\cite{Panteleev_ITI2022}
\begin{equation}
\begin{split}
H_X &= \begin{bmatrix}H_1\boxtimes \mathbb{I}_{n_2}&\mathbb{I}_{m_1}\boxtimes H_2^\top\end{bmatrix}~,\\
H_Z &= \begin{bmatrix}\mathbb{I}_{n_1}\boxtimes H_2&H_1^\top\boxtimes \mathbb{I}_{m_2}\end{bmatrix}~.
\end{split}
\label{eq:lp}
\end{equation}
At the level of the notation this is identical to the HGP code; the difference arises from applying the addition and multiplication operations of the ring of circulants rather than using the $\F_2$-valued parity check matrices \cite{Panteleev_ITI2022}.
The example of Eq.~\eqref{eq:cubic_fsl} is particularly simple, with $H_1 = [\lambda_f]$ and $H_2 = [\lambda_{\overline g}]$.
The LP code thus acts on $L^3(n_1 n_2 + m_1 m_2) = 2L^3$ qubits, as expected.

For comparison, in the HGP code, qubits are defined on the Cartesian product of the coordinates: that is, $\tilde i\equiv(i_1,i_2)$, $\tilde j\equiv(j_1,j_2)$, and $\tilde k\equiv(k_1,k_2)$, each parameterizing a 2D Euclidean plane.
The complete coordinate system $(\tilde i,\tilde j,\tilde k) \equiv(i_1,i_2,j_1,j_2,k_1,k_2)$ is a six-dimensional Euclidean space, in which the HGP code is local.
In contrast, the LP code reduces the coordinate $\tilde i \to (i_1+i_2)\mod L$ and similarly for $\tilde j$ and $\tilde k$, maintaining locality while reducing the number of qubits by a factor of $L^3$, the order of the symmetry group.

Many known CSS stabilizer models, fracton and otherwise, follow the pattern of Eq.~\eqref{eq:cubic_fsl}: for example, the Sierpinski prism model has $f=1+z$ and $g=1+x+y$~\cite{Yoshida_PRB2013}, so is obtained straightforwardly as a HGP.
The checkerboard model, a canonical instance of Type-I fractons, is given by $f=1+x+y+z$ as in Haah's code, but $g = \overline f$, so is representable as a LP.
Interestingly, the honeycomb color code, a topologically ordered model in two dimensions, is produced as a LP with $f=1+x+y$ and $g = \overline f$~\cite{Vijay_PRB2015}.
The color code is equivalent to two copies of the toric code, the prototypical HGP~\cite{Kubica_2015}.

\subsubsection{X-cube model}\label{sec:x_cube_model}

Another canonical instance of fracton order is the $X$-cube model, which is Type I, realizing lineons and planons \cite{Vijay_PRB2016}.
The checkerboard model shown to be an LP in the previous section is equivalent to two copies of the $X$-cube model~\cite{Shirley19}. 
The generating map of this code is
\begin{equation}
    \sigma_{X\text{-cube}}=\left(\begin{array}{ccc}
    \overline g \overline h & 0 & 0\\
    \overline f \overline h & 0 & 0\\
    \overline f \overline g & 0 & 0\\
    \hline
    0 & f & 0\\
    0 & g & g\\
    0 & 0 & h\\
\end{array}\right),
\end{equation}
where $f=1+x$, $g=1+y$, and $h=1+z$.
That is, $H_X=\begin{bmatrix}\lambda_{\overline g \overline h}&\lambda_{\overline f \overline h}&\lambda_{\overline f \overline g}\end{bmatrix}$ and $H_Z=\begin{bmatrix}\lambda_f&\lambda_g&0\\0&\lambda_g&\lambda_h\end{bmatrix}$.
This model generalizes the prior example of Haah's code in several ways.
First, there are two independent types of $Z$-type stabilizer and only a single $X$-type stabilizer, the eponymous $X$-cube; in addition, the unit cell contains three qubits.

The latter observation suggests that the $X$-cube model is most naturally obtained as a \emph{threefold} product code.
For such generalized products, it is convenient to employ the language of homology.
Recalling that codewords are 0 eigenstates of the parity check matrix, a classical code can be thought of as a chain complex $\C_1 \xrightarrow{\partial_1} \C_0$, whose $1$-chains are its parity checks and $0$-cycles are codewords; thus $\partial_1 = H^\top$.
A quantum CSS code is similarly represented as $\C_2 \xrightarrow{\partial_2} \C_1 \xrightarrow{\partial_1} \C_0$, where now $\C_2$ is the group of $Z$ stabilizers, $\C_1$ the codewords, and $\C_0$ the $X$ stabilizers.
Accordingly, $\partial_2 = H_Z^\top$ and $\partial_1 = H_X$.
These evidently satisfy the boundary map property $\partial_2 \partial_1 = 0$ by the definition of a stabilizer code.

Up to a conventional difference in transposes, the HGP is seen to be the tensor product of chain complexes associated with the seed codes.
This generalizes to \emph{homological product codes} obtained by the multifold tensor product of chain complexes~\cite{Bravyi_ACM2014}.
The longer chain complexes obtained in this way feature meta-checks which provide classical error correction for the syndrome measurement outcomes~\cite{Quintavalle_PRX2021}.
In order to construct the $X$-cube model, we consider instead a particular multifold product which is not the tensor product of chain complexes: given inputs $\{\C_1^1 \xrightarrow{\partial_1} \C_0^1,\C_1^2 \xrightarrow{\partial_2} \C_0^2,\C_1^3 \xrightarrow{\partial_3} \C_0^3 \}$ with $\partial_1 = H_1^\top$, $\partial_2 = H_2^\top$, and $\partial_3 = H_3^\top$, we write a 2-complex $\C_2^q \xrightarrow{\partial_2^q} \C_1^q \xrightarrow{\partial_1^q} \C_0^q$, whose boundary maps are defined as
\begin{equation}
\begin{gathered}
\partial_2^q = \begin{bmatrix} H_1^\top\otimes\I_2\otimes\I_3 & H_1^\top\otimes\I_2\otimes\I_3 & 0 \\ \I_1\otimes H_2^\top\otimes\I_3 & 0 & \I_1\otimes H_2^\top\otimes\I_3 \\ 0 & \I_1\otimes\I_2\otimes H_3^\top & \I_1\otimes\I_2\otimes H_3^\top \end{bmatrix},\\
\partial_1^q = \begin{bmatrix} \I_1\otimes H_2^\top\otimes H_3^\top & H_1^\top\otimes\I_2\otimes H_3^\top & H_1^\top\otimes H_2^\top\otimes\I_3 \end{bmatrix}.
\end{gathered}
\end{equation}
In fact, similar boundary maps have been discussed in the setting of the $X$-cube model, but in a different context~\cite{Radicevic_Arxiv2020}.
The vector spaces over $\F_2$ are
\begin{equation}
\begin{gathered}
\C_2^q = (\C_1^1 \otimes \C_1^2 \otimes \C_1^3) \oplus (\C_1^1 \otimes \C_1^2 \otimes \C_1^3) \oplus (\C_1^1 \otimes \C_1^2 \otimes \C_1^3)~,\\
\C_1^q = (\C_0^1\otimes\C_1^2\otimes\C_1^3) \oplus (\C_1^1\otimes\C_0^2\otimes\C_1^3) \oplus (\C_1^1\otimes\C_1^2\otimes\C_0^3)~,\\
\C_0^q=\C_0^1\otimes \C_0^2\otimes \C_0^3~.
\end{gathered}
\end{equation}

This definition of a threefold product code obtains the $X$-cube model as $\partial_2^q = H_Z^\top$ and $\partial_1^q = H_X$ when $H_1 = [\lambda_f]$, $H_2 = [\lambda_g]$, $H_3 = [\lambda_h]$, each defined on a cycle $C_L$.
Evidently the correct number of qubits $n_q = 3L^3$ arises without performing any symmetry reduction; that is, we attach a different label $\{x,y,z\}$ to each cycle and do not take a quotient on the input codes.
In this way the model is found to be essentially a threefold generalization of the HGP code (though such a definition is not unique) and is thus distinguished from the LP construction required for Haah's code.
Moreover, we reproduce that the $X$-cube model does not have meta-checks, in contrast with, e.g., the three-dimensional toric code, obtained as the standard homological product of the same classical seeds~\cite{Quintavalle_PRX2021}.

\end{document}